\documentclass[nofootinbib,reprint,amsmath,amssymb,aps]{revtex4-1}
\usepackage{graphicx}
\usepackage[utf8,latin1]{inputenc}
\usepackage{dcolumn}
\usepackage{bm}
\usepackage{hyperref}
\usepackage{amsmath}
\newcommand{\qm}[1]{``#1''}
\usepackage{hyperref}
\hypersetup{colorlinks, linkcolor={red},citecolor={blue},urlcolor={blue}}  

\newcommand\RS{{\mathrm{R_\star}}}
\newcommand{\vdf}[1]{\textcolor{red}{#1}}

\begin{document}

\title[Detection of chaos in the general relativistic Poynting-Robertson effect: Kerr equatorial plane]{Detection of chaos in the general relativistic\\ Poynting-Robertson effect: Kerr equatorial plane}

\author{Vittorio De Falco$^{1}$}\email{vittorio.defalco@ibf.cnr.it}\ \email{vittorio.defalco@physics.cz}
\author{William Borrelli$^{2}$} \email{william.borrelli@unicatt.it}

\affiliation{$^1$Department of Mathematics and Applications \qm{R. Caccioppoli}, University of Naples Federico II, Via Cintia, 80126 Naples, Italy, \\
$^2$Dipartimento di Matematica e Fisica, Universit\`a Cattolica del Sacro Cuore, Via dei Musei 41, Brescia, Italy.}

\date{\today}

\begin{abstract}
The general relativistic Poynting-Robertson effect is a dissipative and non-linear dynamical system 
obtained by perturbing through radiation processes the geodesic motion of test particles orbiting around a spinning compact object, described by the Kerr metric. Using the Melnikov method we find that, in a suitable range of parameters, chaotic behavior is present in the motion of a test particle driven by the Poynting-Robertson effect in the Kerr equatorial plane.
\end{abstract}

\maketitle

\section{Introduction}
\label{sec:intro}
Chaos is a widespread feature in many physical non-linear dynamical systems. A chaotic system hides behind the visible randomness of the complex dynamics, some underlying rich mathematical structures, such as: constant feedback loops, self-similarities, fractals, and self-organization \cite{Wiggins1988,Ott2002}. Although an universally accepted formal definition of chaos does not exist, the one due to Robert L. Devaney is widely accepted, and it is based on the following three proprieties \cite{Devaney2018}: 
\begin{itemize}
\item $(1)$ \emph{sensitive dependence on initial conditions}, i.e., tiny perturbations on the initial conditions leads to significantly different future behaviors, 
\item $(2)$ \emph{topologically mixing}, i.e., any given region or open set of the phase space eventually overlaps with any other given region in the phase space;
\item $(3)$ \emph{presence of a dense set of periodic orbits}, i.e., every point in the dynamical real space is approached arbitrarily close by periodic orbits. 
\end{itemize}

General Relativity (GR), being a non-linear theory, can potentially exhibit chaotic phenomena \cite{Hobill1994}. The studies on chaos in GR can be mainly divided in two branches: (1) problems of geodesic/non-geodesic motion of a particle in a given gravitational field; (2) evolution of cosmological models. Regarding works on the first class, it is important to mention: the motion in spaces with negative curvature \cite{Arnold1989}, motion around two fixed black holes (BHs) \cite{Contopoulos1990,Contopoulos1991,Hobill1994}, relativistic restricted three-body problem \cite{Wanex2002}, Schwarzschild BH affected by high-frequency periodic perturbations \cite{Bombelli1992}, spinning particle motion around a Kerr and Schwarzschild BH \cite{Suzuki1997,Lukes2018}, gravitational waves from spinning compact binaries \cite{Cornish2001,Cornish2002,Cornish2003}. Moreover, studies on chaos in cosmology include:  the model of Belinski-Khalatnikov-Lifshitz dealing with the dynamic evolution of the universe near the initial gravitational singularity, described by an anisotropic, chaotic solution of the Einstein field equation of gravitation. \cite{Belinskij1970a,Belinskij1970b}, Bianchi IX (\qm{mixmaster universe}) \cite{Barrow1989,Burd1991,Contopoulos1999}, Friedmann-Robertson-Walker (FRW) plus a massive scalar field \cite{Calzetta1993}, and the non-linear interaction among dark matter, dark energy, normal matter, and radiation on the FRW spacetime \cite{Aydiner2016}. 

In high-energy astrophysics, dealing with electromagnetic radiation processes around compact objects, like neutron stars (NSs) or BHs, relatively small-sized test particles can drastically depart from their geodesic motion. The gravitational pull is contrasted by the radiation pressure, and in the process of absorption and reemission of radiation from the test particle an additional radiation torque appears, acting as a drag force opposite to the test particle orbital motion \cite{Poynting1903,Robertson1937}. \emph{This is the so-called Poynting-Robertson (PR) effect, which configures thus as a dissipative non-linear dynamical system efficiently removing energy and angular momentum from the affected test particle}. There are several models of the general relativistic PR effect in Kerr and also other spacetimes from the two dimensional (2D) \cite{Bini2009,Bini2011,Bini2015} to the three dimensional (3D) formulations \cite{DeFalco20183D,Bakala2019,Wielgus2019,DeFalco2020,DeFalcoTESI}. They all exhibit the existence of a critical hypersurface, a region where gravitational and radiation forces balance and the test particle moves on it stably \cite{Bini2011,DeFalco2019ST,DeFalco2020sum}. 

Here, we focus our attention on the \emph{general relativistic PR effect in the equatorial plane around a Kerr compact object}. To search for chaotic behavior, we employ the \emph{Melnikov method} \cite{Guckenheimer2002,Wiggins1988}, which is an independent diagnostic procedure, complementary to other numerical and analytical methods \cite{Tabor1989,Ott2002,Guckenheimer2002}. Its strength relies on the fact, that \emph{it requires only the knowledge of few elements without having any insight into the solution of the perturbed dynamics, i.e.: (1) invariant subsets in the phase space of the unperturbed dynamics (homoclinic orbits); (2) explicit expression of the perturbations}. 

The paper is organized as follows: in Sec. \ref{sec:PRmodel} we briefly recall the general relativistic PR effect model, underlining also how to derive its dissipative perturbations; in Sec. \ref{sec:HO} the homoclinic orbits in the equatorial plane of the Kerr spacetime are described; in Sec. \ref{sec:MI} we apply the Melnikov method to the general relativistic PR effect; in Sec. \ref{sec:end} we discuss our results and draw the conclusions. 

\section{General relativistic Poynting-Robertson effect in the equatorial plane of the Kerr metric}
\label{sec:PRmodel}
The general relativistic PR effect in the Kerr metric describes the motion of a test particle influenced by the gravitational field, the radiation pressure and the radiation drag force. The radiation field is modeled by photons stemming out from a spherical and rigidly rotating radiation source, which permits to calculate their impact parameter. They move along null geodesics of the Kerr metric and hit the test particle at each time instant, modifying thus its timelike geodesic trajectory (see Sec. \ref{sec:model}). We underline the ranges of the model parameters, which will be useful in Sec. \ref{sec:MI}. This model can be recast in Hamiltonian form, convenient to extract the dissipative PR perturbations (see Sec. \ref{sec:HF}). Finally, we discuss some a-priori indications of possible chaotic behaviour in the general relativistic PR effect (see Sec. \ref{sec:ICB}). 

\subsection{The model}
\label{sec:model}
We consider a central compact object, whose outside spacetime is described by the Kerr metric with signature $(-,+,+,+)$. In geometrical units ($c = G = 1$), the line element of the Kerr spacetime, $ds^2=g_{\alpha\beta}dx^\alpha dx^\beta$, in Boyer-Lindquist coordinates, parameterized by mass $M$ (set equal to unity, $M=1$) and spin $a$, settled in the equatorial plane $\theta=\pi/2$, reads as
\begin{equation}\label{kerr_metric}
\begin{aligned}
 \mathrm{d}s^2 &= \left(\frac{2}{r}-1\right)\mathrm{d}t^2 
  - \frac{4a}{r}\mathrm{d}t\, \mathrm{d}\varphi + \frac{r^2}{\Delta}\, \mathrm{d}r^2 
  + \rho \mathrm{d}\varphi^2, 
\end{aligned}
\end{equation}
where $\Delta \equiv r^{2} - 2r + a^{2}$, and $\rho  \equiv r^2+a^2+2a^2/r$. We introduce the zero angular momentum observers (ZAMOs), whose adapted orthonormal frame is given by\footnote{The hat over the indices indicates that the corresponding vector or tensor quantity is calculated in the ZAMO frame.} \cite{Bini2009,Bini2011}
\begin{eqnarray} \label{eq:zamoframes}
&&\boldsymbol{e_{\hat t}}\equiv\boldsymbol{n}=\frac{\boldsymbol{\partial_t}-N^{\varphi}\boldsymbol{\partial_\varphi}}{N},\ \boldsymbol{e_{\hat r}}=\frac{\boldsymbol{\partial_r}}{\sqrt{g_{rr}}},\ \boldsymbol{e_{\hat \varphi}}=\frac{\boldsymbol{\partial_\varphi}}{\sqrt{g_{\varphi \varphi }}},
\end{eqnarray}
where $\left\{\boldsymbol{\partial_t},\ \boldsymbol{\partial_r},\ \boldsymbol{\partial_\varphi}\right\}$ is the orthonormal frame adapted to the static observer at infinity, $N=(-g^{tt})^{-1/2}$ is the time lapse function and $N^{\varphi}=g_{t\varphi}/g_{\varphi\varphi}$ the spatial shift vector field, whose explicit expressions are \cite{Bini2011}
\begin{equation}
N=\sqrt{\frac{\Delta}{\rho}},\qquad N^\varphi=-\frac{2a}{r\Delta}.
\end{equation}
 
The radiation field is constituted by a coherent flux of photons traveling along null geodesics in the Kerr geometry. The related stress-energy tensor is \cite{Bini2009,Bini2011,DeFalco20183D,Bakala2019}
\begin{equation}\label{STE}
T^{\mu\nu}=\Phi^2 k^\mu k^\nu\,,\qquad k^\mu k_\mu=0,\qquad k^\mu \nabla_\mu k^\nu=0,
\end{equation}
where $\boldsymbol{k}$ is the photon four-momentum field, and $\Phi$ is a parameter linked to the radiation field intensity, whose explicit expression is given by \cite{Bini2011} \footnote{The radial radiation function $\mathcal{R}_{\rm rad}(r)$ can be equivalently written as $\mathcal{R}_{\rm rad}(r)=(r^2+a^2-ab)^2-\Delta(a-b)^2$ \cite{DeFalco20183D,Bakala2019}.}
\begin{equation}\label{INT_PAR}
\Phi^2=\frac{\Phi_0^2}{\sqrt{\mathcal{R}_{\rm rad}(r)}},\qquad \mathcal{R}_{\rm rad}(r)=rN|b\tan\beta|,
\end{equation}
where $\Phi_0$ is $\Phi$ evaluated at the emitting surface. Splitting $\boldsymbol{k}$ with respect to the ZAMOs, we obtain \cite{Bini2009,Bini2011}
\begin{eqnarray}
&&\boldsymbol{k}=E(n)[\boldsymbol{n}+\boldsymbol{\hat{\nu}}],\qquad \boldsymbol{\hat{\nu}}=\sin\beta\ \boldsymbol{e_{\hat r}}+\cos\beta\ \boldsymbol{e_{\hat\varphi}},
\end{eqnarray}
where $\boldsymbol{\hat{\nu}},\ \beta,\ E(n)=E_p(1+bN^\varphi)/N$ 
with $E_p=-k_t$ is the conserved photon energy along its trajectory are respectively the photon spatial unit relative velocity, the angle in the azimuthal direction, and the photon energy, where all quantities are measured in the ZAMO frame \cite{Bini2009,Bini2011},
 The radiation field is governed by the impact parameter $b$, associated with the emission angle $\beta$. 

The photons of the radiation field are emitted from a spherical surface having radius $\RS$ centered at the origin of the Boyer-Lindquist coordinates, and rigidly rotating with angular velocity $\Omega_{\mathrm{\star}}\ge0$. Defined the event horizon $R_{\rm H}=1+\sqrt{1-a^2}$ and the static limit $R_{\rm SL}=2$ radii in the equatorial plane, we have that $R_\star\in(R_{\rm H}(a),\bar{R}_\star]$, where $\bar{R}_\star<\infty$. Once $R_\star$ has been chosen, we want that $\Omega_\star\in[\Omega_{\rm min},\Omega_{\rm max}]=[\Omega_-,\Omega_+]\cap[0,\Omega_+]$, where \cite{Bakala2019}
\begin{equation}
\Omega_\pm=\frac{-g_{t\varphi}\pm\sqrt{g_{t\varphi}^2-g_{\varphi\varphi}g_{tt}}}{g_{\varphi\varphi}}.
\end{equation}
The photon impact parameter is given by \cite{Bakala2019}
\begin{eqnarray} 
&&b=-\left[\frac{\mathrm{g_{t\varphi}}+\mathrm{g_{\varphi\varphi}}\Omega_{\star} }{\mathrm{g_{tt}}+\mathrm{g_{t\varphi}} \Omega_{\star}}\right]_{r=\RS},\label{kerr_impact_parameter}
\end{eqnarray}
which in these premises ranges in $[b_{\rm min},b_{\rm max}]\subseteq\mathbb{R}$ (see Fig. 2 in Ref. \cite{Bakala2019}, for more details). The related photon angle in the ZAMO frame is \cite{Bakala2019}
\begin{equation} \label{ANG1}
\cos\beta=\frac{b N}{\sqrt{g_{\varphi\varphi}}(1+b N^\varphi)}, 
\end{equation}
where $\beta\in[0,2\pi]$. For $\sin\beta>0$ ($\sin\beta<0$) we are considering outgoing (ingoing) photons, see Ref. \cite{Bini2011}. 

A test particle moves with a timelike four-velocity $\boldsymbol{U}$ and a spatial three-velocity with respect to the ZAMOs, $\boldsymbol{\nu}(U,n)$, which both read as \cite{Bini2009,Bini2011,DeFalco20183D,Bakala2019}
\begin{eqnarray} \label{eq:veloci}
&&\boldsymbol{U}=\gamma[\boldsymbol{n}+\boldsymbol{\nu}],\qquad \boldsymbol{\nu}=\nu(\sin\alpha\boldsymbol{e_{\hat r}}+\cos\alpha \boldsymbol{e_{\hat\varphi}}),
\end{eqnarray}
where $\gamma=1/\sqrt{1-||\boldsymbol{\nu}||^2}$ is the Lorentz factor, $\nu=||\boldsymbol{\nu}||$ is the magnitude of the test particle spatial velocity $\boldsymbol{\nu}(U,n)$, $\alpha$ is the azimuthal angle of the vector $\boldsymbol{\nu}$ measured clockwise from the positive $\boldsymbol{\hat\varphi}$ direction in the $\boldsymbol{\hat{r}}-\boldsymbol{\hat{\varphi}}$ tangent plane in the ZAMO frame. The energy absorbed by the test particle affected by the incoming photon is $E(U)=-k_\mu U^\mu$, which can be related to the photon energy $E(n)$ in the ZAMO frame through \cite{Bini2009,Bini2011,DeFalco20183D,Bakala2019}
\begin{equation} \label{enepart}
E(U)=\gamma E(n)[1-\nu\sin\psi\cos(\alpha-\beta)].
\end{equation}

We assume that the radiation test particle interaction occurs through Thomson scattering, characterized by a constant momentum-transfer cross section $\sigma$, independent from direction and frequency of the radiation field. The radiation force is given by \cite{Bini2009,Bini2011,DeFalco20183D,Bakala2019}
\begin{equation} \label{radforce}
{\mathcal F}_{\rm (rad)}(U)^{\hat \alpha}=\sigma \, [\Phi E(U)]^2\, \hat{\mathcal V}(k,U)^{\hat\alpha}\,.
\end{equation}
where the term $\tilde{\sigma}[\Phi E(U)]^2$ reads as \cite{Bini2009,Bini2011,DeFalco20183D,Bakala2019} 
\begin{eqnarray} 
\frac{\tilde{\sigma}[\Phi E(U)]^2}{A}&=&\frac{\gamma^2(1+bN^\varphi)^2\mathbb{A}^2}{N^2\sqrt{\mathcal{R}_{\rm rad}(r)}},\label{eq:FACT}\\
\mathbb{A}&=&\left[\gamma-\frac{p_r}{\sqrt{g_{rr}}}\sin\beta-\frac{p_\varphi}{\sqrt{g_{\varphi\varphi}}}\cos\beta\right].\label{eq:AA}
\end{eqnarray}
The term $A=\tilde{\sigma}[\Phi_0 E_p]^2$ is the luminosity parameter, which can be equivalently written as $A=L/L_{\rm Edd}\in[0,1]$, with $L$ the emitted luminosity at infinity and $L_{\rm Edd}$ the Eddington luminosity. We have that $\tilde{\sigma}=\sigma/m$, where $m$ is the test particle mass, which for easing the notations we set equal to unity, $m=1$. The terms $\hat{\mathcal V}(k,U)^{\hat\alpha}$ are the radiation field components, which are \cite{Bini2009,Bini2011,DeFalco20183D,Bakala2019}
\begin{eqnarray}
\hat{\mathcal{V}}_{\hat r}&=&\frac{1}{\mathbb{A}}\left[\sin\beta-\frac{p_r}{\sqrt{g_{rr}}}\mathbb{A}\right],\label{eq:Vr}\\
\hat{\mathcal{V}}_{\hat \varphi}&=&\frac{1}{\mathbb{A}}\left[\cos\beta-\frac{p_\varphi}{\sqrt{g_{\varphi\varphi}}}\mathbb{A}\right],\label{eq:Vf}\\
\hat{\mathcal{V}}_{\hat t}&=&\frac{1}{\mathbb{A}}\left[1-\gamma\mathbb{A}\right],\label{eq:Vt}
\end{eqnarray}

\subsection{Hamiltonian formulation}
\label{sec:HF}
The general relativistic PR effect in the Lagrangian formalism has been already treated in \cite{DeFalco2018,DeFalco2019,DeFalco2019VE,DeFalco2020sum}, and we now pass to its Hamiltonian formulation. In the geodesic case, we consider the mass shell constraint $g^{\alpha\beta}p_\alpha p_\beta=-1$, where the momentum $p_\alpha$ is canonically conjugate to $x^\alpha$ through the \emph{Legendre transform} $p_\alpha=g_{\alpha\beta}\dot{x}^\beta$. Here the dot stands for the derivative with respect to the affine parameter $\tau$. Therefore, the Hamiltonian is $\mathcal{H}(\boldsymbol{p},\boldsymbol{x})=g^{\alpha\beta}p_\alpha p_\beta/2$
and, the Hamilton equations are
\begin{equation} \label{eq:HEqs}
\dot{x}^\mu=\frac{\partial \mathcal{H}}{\partial p_\mu},\qquad \dot{p}_\mu=-\frac{\partial \mathcal{H}}{\partial x^\mu}.
\end{equation}
Such formulation can be also extended to a dissipative system, where the perturbations $\boldsymbol{f}(\boldsymbol{p},\boldsymbol{x})=(f_1^\mu,f_{2,\mu})$ are not of Hamiltonian type, therefore Eqs. (\ref{eq:HEqs}) become
\begin{equation}\label{eq:Hmod}
\begin{aligned}
&\dot{x}^\mu=\frac{\partial \mathcal{H}}{\partial p_\mu}+\epsilon f_1^\mu,\qquad \dot{p}_\mu=-\frac{\partial \mathcal{H}}{\partial x^\mu}+\epsilon f_{2,\mu},
\end{aligned}
\end{equation}
where $\epsilon\ll1$ is a small parameter.

\subsubsection{General relativistic PR perturbations}
\label{sec:PRinHF}
The test particle velocity components are \cite{Bini2009,Bini2011}
\begin{eqnarray}
&&U^{\hat r}\equiv\frac{dr}{d\tau}=\frac{\gamma\nu\sin\alpha}{\sqrt{g_{rr}}}, \label{EoM4}\\
&&U^{\hat\varphi}\equiv\frac{d\varphi}{d\tau}=\frac{\gamma\nu\cos\alpha}{\sqrt{g_{\varphi\varphi}}}-\frac{\gamma N^\varphi}{N},\label{EoM6}\\
&&U^{\hat t}\equiv \frac{dt}{d\tau}=\frac{\gamma}{N},\label{time}
\end{eqnarray}
where $\tau$ is the affine parameter (proper time) along the test particle trajectory, see Eqs. (\ref{eq:veloci}). In the PR effect case, the conjugate momenta $p_\mu$ to the $x^\mu=(t,r,\varphi)$ are
\begin{eqnarray}
&&\frac{p_r}{\sqrt{g_{rr}}}=\gamma\nu\sin\alpha, \quad
\frac{p_\varphi}{\sqrt{g_{\varphi\varphi}}}=\gamma\nu\cos\alpha,\quad p_t=\frac{\gamma}{N}. \label{eq:prf}
\end{eqnarray}
In such formalism, we have that $\nu$ and $\gamma$ read as
\begin{equation} \label{eq:velga}
\nu=\sqrt{\frac{\left(\frac{p_r^2}{g_{rr}}+\frac{p_\varphi^2}{g_{\varphi\varphi}}\right)}{1+\left(\frac{p_r^2}{g_{rr}}+\frac{p_\varphi^2}{g_{\varphi\varphi}}\right)}},\quad \gamma=\sqrt{1+\frac{p_r^2}{g_{rr}}+\frac{p_\varphi^2}{g_{\varphi\varphi}}}.
\end{equation}
Using the radiation force components (\ref{eq:Vr}) -- (\ref{eq:Vt}), we obtain $\tilde{F}_\mu=\tilde{\sigma}[\Phi E(U)]^2\hat{\mathcal{V}}_{\hat \mu}/A$,
where $\epsilon=A\equiv L/L_{\rm Edd}\ll1$, namely low luminosities. The PR dissipative perturbations are $(f^\mu_1,f_{2,\mu})=(0,\tilde{F}_\mu)$. We note that $f^\mu_1=0$, because the radiation field, including radiation pressure and PR drag force, affects only the accelerations and not the velocity components, see Refs. \cite{Bini2011} for more details.

\subsection{A-priori indications of chaotic behavior}
\label{sec:ICB}
The main motivations for the present study are explained in this section. During the investigation of the general relativistic PR effect, a series of a-priori indications of possible chaotic dynamics can be found:
\begin{itemize}
\item it is a dissipative and non-linear dynamical system in GR \cite{DeFalco2019VE}, which makes the Kerr geodesic motion not \emph{integrable} \cite{Strogatz1994,Mori2013}; 

\item it has been analytically and numerically confirmed that such effect generally behaves as a \emph{forced harmonic oscillator} \cite{Bini2009,Bini2011} endowed with a non-linear driven force (close to a Duffing oscillator), responsible to potentially create \emph{resonance effects} \cite{Tabor1989};  

\item a test particle under the general relativistic PR effect can end its motion either on the critical hypersurface or escaping at infinity. It has been already formally proved that the critical hypersurface behaves as a \emph{stable attractor} \cite{DeFalco2019ST}, and the same holds also for the spatial infinity (never returning back); 

\item as proved in \cite{Bini2009}, such effect admits positive \emph{Lyapunov exponents}, which measure the mean rate of exponential separation of neighboring trajectories \cite{Tabor1989}. This is an useful index indicating that a dynamical system shows \emph{sensitive dependence on the initial conditions}. This propriety has been further confirmed by numerical simulations.
\end{itemize}

\section{Homoclinic orbits}
\label{sec:HO}
The notion of homoclinic orbits for a dynamical system is based on the research of \emph{recurrent invariant sets} $\Lambda$ \cite{Wiggins1988,Guckenheimer2002}, such as fixed points, periodic orbits, or $n$-dimensional invariant tori. The set of all trajectories approaching an invariant set $\Lambda$ asymptotically in the infinite future (past) is a submanifold of the phase space termed \emph{stable (unstable) manifold of $\Lambda$}, usually indicated by $W^s(\Lambda)$ ($W^u(\Lambda)$). An invariant set $\Lambda$ possessing both stable and unstable manifolds is called \emph{hyperbolic}\footnote{The given definition has a clear dynamical meaning. However, the rigorous definition of an hyperbolic point $p$ for a $C^1$ vector field $\boldsymbol{F}: \mathbb{R}^n \to \mathbb{R}^n$ is the following: $p$ is a critical point for $\boldsymbol{F}$, i.e., $\boldsymbol{F}(p) = 0$, and the Jacobian matrix of $\boldsymbol{F}$ at $p$, $\boldsymbol{J}=(\nabla \boldsymbol{F})(p)$, has no eigenvalues with zero real parts \cite{Wiggins1988,Guckenheimer2002}. The stable (unstable) manifold of $p$ consists of points $q$ such that $\phi_t(q)\to p$ as $t\to+\infty$ ($t\to-\infty$), where $\phi_t$ is the \emph{flow} associated with $\boldsymbol{F}$.}. 

A trajectory is defined to be homoclinic to a hyperbolic invariant set $\Lambda$ if it approaches $\Lambda$ in the infinite future as in the infinite past, i.e., $W^u(\Lambda)\cap W^s(\Lambda)\cap\Lambda\neq\emptyset$ \cite{Wiggins1988,Guckenheimer2002}. \emph{Therefore, for determining the class of the homoclinic orbits of a dynamical system we need to identify the intersections of their stable and unstable manifolds on their hyperbolic invariant sets}.

\subsection{Homoclinic orbits in the equatorial plane of Kerr spacetime}
\label{sec:Levin}
We consider the following dynamical system represented by the motion of a timelike test particle governed only by gravity and no other perturbing effects in the equatorial plane of the Kerr metric \cite{Misner1973,Levin2009}

\begin{eqnarray}
&&\dot{t}=\frac{r\rho E-2aL_z}{r\Delta},\quad \dot{r}=\pm\frac{\sqrt{R(r)}}{r^2},\quad  \dot{\varphi}=\dot{\varphi}(a,r), \label{eq:r}
\end{eqnarray}
where $\dot{\varphi}(a,r)=[2aE+L_z(r-2)]/(r\Delta)$, \footnote{It is possible to factorize $R(r)=-(1-E^2)r^4+2r^3-[a^2(1-E^2)+L_z^2]r^2+2(aE-L_z)^2r$ as reported in Eq. (\ref{eq:FR}).}
\begin{equation} \label{eq:FR}
\begin{aligned}
&R(r)=-(1-E^2)r(r-r_u)^2(r-r_a), 
\end{aligned}
\end{equation}
$E=-p_t$, and $L_z=p_\varphi$ are respectively the energy and angular momentum with respect to the $\boldsymbol{z}$-axis (orthogonal to the equatorial plane) conserved along the test particle trajectory, and $r_u$ and $r_a$ are respectively the periastron and apastron radii of the homoclinic orbit. Throughout the paper the signs $\pm$ refers to prograde and retrograde orbits, respectively.

The invariant sets are the circular orbits (defined by the conditions $R(r)=0$ and $dR(r)/dr=0$), while the hyperbolic invariant sets coincide with the unstable circular orbits (defined by circular orbit condition, and $d^2R(r)/dr^2<0$, which corresponds to the maximum of $dR/dr=0$). Among these trajectories, the homoclinic orbits are the unstable circular orbits energetically bounded ($E<1$) \cite{Levin2009}, that we describe through the periastron and apastron radii $(r_p,r_a)$. \emph{Homoclinic orbits are in a one-to-one correspondence with bound energy values $E<1$, and therefore constitute a one-parameter family specified by the (periastron) radius $r_u=r_p$}. 
\begin{figure*}[t!]
\centering
\leavevmode
\hbox{\includegraphics[scale=0.22]{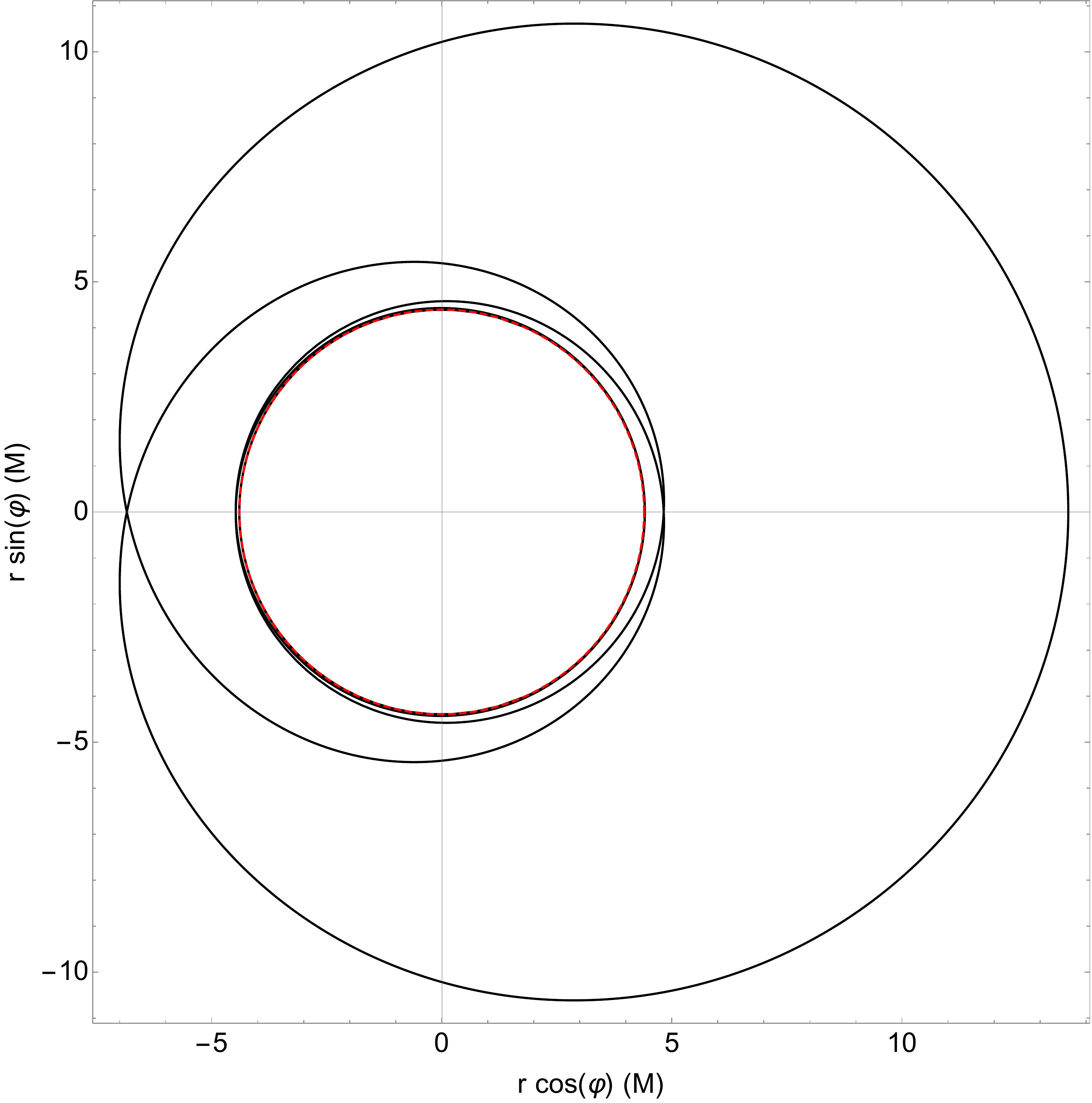}
\hspace{1cm}
\includegraphics[trim=0.5cm 0cm 1cm 0.5cm,scale=0.36]{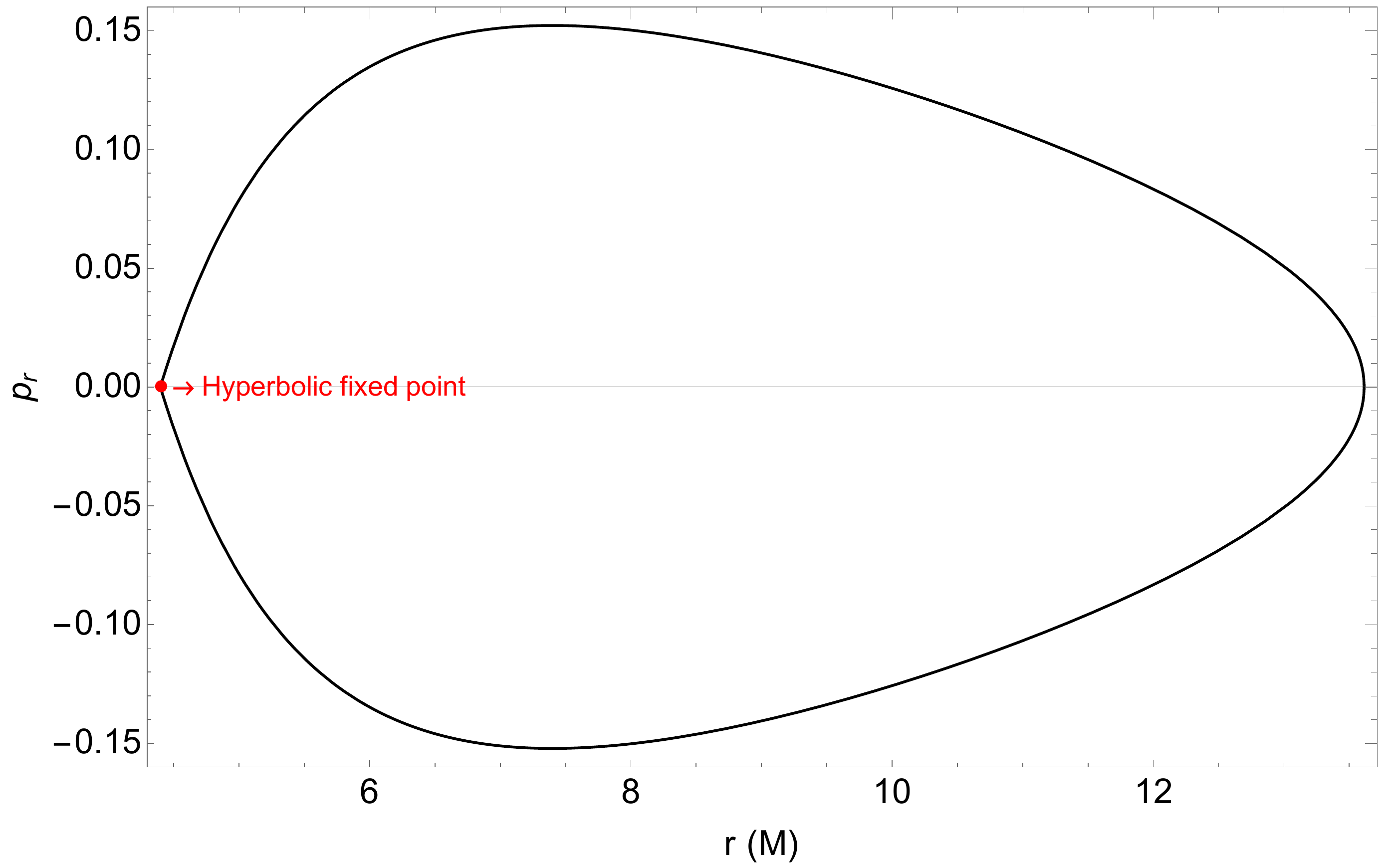}}
\caption{Homoclinic orbit (left panel) and its phase portrait (right panel), for $a=0.1$, $r_u=4.40$. The other parameters are $r_a=13.61$, $E=0.95$, and $L_z=3.52$. The red dashed line represents the circular orbit of radius $r_u$ centered in $(0,0)$ towards which the homoclinic orbit moves. The hyperbolic fixed red point in the phase space has coordinates $(r,p_r)=(r_u,0)$.}
\label{fig:Plot1}
\end{figure*}

\begin{figure}[h!]
\includegraphics[trim=0.5cm 0cm 0cm 1.5cm, scale=0.29]{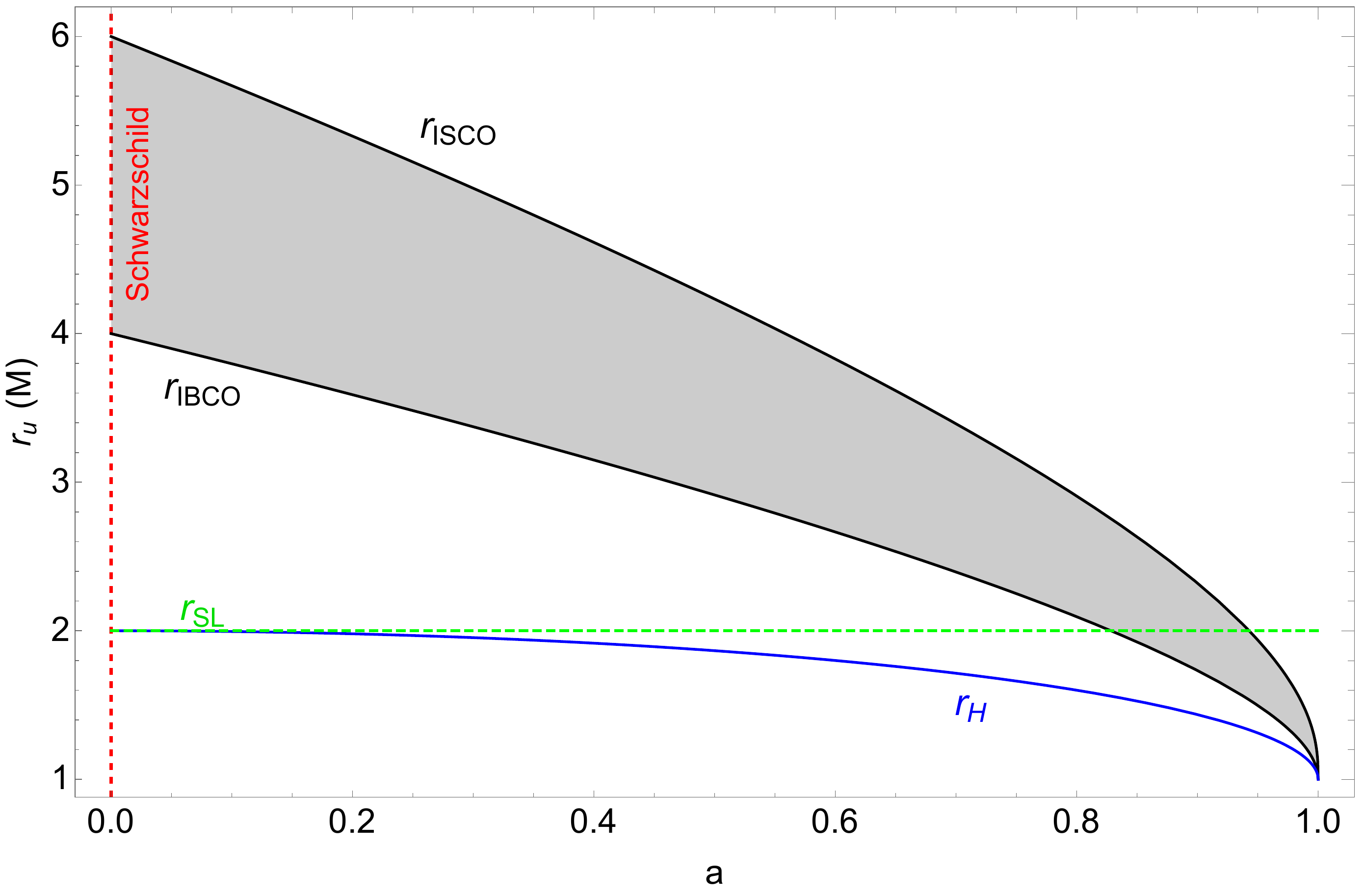}
\caption{The shaded area, delimited by IBCO and ISCO radii, is where $r_u$ can range in terms of the spin $a$. The vertical dashed red line represent the Schwarzschild limit $(a=0)$, the horizontal dashed green line is the static limit radius, and the continuous blue line is the event horizon radius.}
\label{fig:Fig3}
\end{figure}

The one-parameter family of homoclinic orbits in the equatorial plane of the Kerr spacetime $\mathcal{O}^{\rm hc}(r_u)$, see Fig. \ref{fig:Plot1} as an example\footnote{To plot the homoclinic orbit in the equatorial plane of Kerr metric, we use Eq. (26c) in Ref. \cite{Levin2009} for describing the azimuthal coordinate $\varphi$, while the radial coordinate $r$ ranges in $[r_u,r_a]$.}, is characterized by \cite{Levin2009}
\begin{eqnarray}
E&=&\frac{r_u^{3/2}-2r_u^{1/2}\pm a}{r_u^{3/4}\sqrt{r_u^{3/2}-3r_u^{1/2}\pm 2a}}<1,\label{eq:enpart}\\
L_z&=&\frac{r_u^{2}\mp2ar_u^{1/2}+ a^2}{r_u^{3/4}\sqrt{r_u^{3/2}-3r_u^{1/2}\pm 2a}},\label{eq:ampart}\\
r_a&=&\frac{2(aE-L_z)^2}{r_u^2(1-E^2)}\equiv\frac{2r_u(a\mp\sqrt{r_u})^2}{r_u^2-4r_u\pm4a\sqrt{r_u}-a^2}, \label{eq:ra}
\end{eqnarray}
where $r_u$ ranges between the innermost bound circular orbit (IBCO), and the innermost stable circular orbit (ISCO), i.e., $r_u\in[r_{\rm IBCO},r_{\rm ISCO}]$ (see Fig. \ref{fig:Fig3}), with
\begin{eqnarray}
r_{\rm IBCO}&=&2\mp a+2\sqrt{1\mp a},\\
r_{\rm ISCO}&=&3+Z_2\mp\sqrt{(3-Z_1)(3+Z_1+2Z_2)},\\
Z_1&=&1+{}^3\sqrt{1-a^2}\left[{}^3\sqrt{1+a}+{}^3\sqrt{1-a}\right],\\
Z_2&=&\sqrt{3a^2+Z_1^2}.
\end{eqnarray}

\section{Melnikov integral}
\label{sec:MI}
The Melnikov method is a powerful mathematical tool to identify the occurrence of chaos in 2D and even higher-dimensional dynamical systems affected by Hamiltonian periodic or non-Hamiltonian perturbations \cite{Wiggins1988,Guckenheimer2002,Holmes1982V1,Holmes1982V2}.

\begin{figure}[h!]
\includegraphics[scale=0.33]{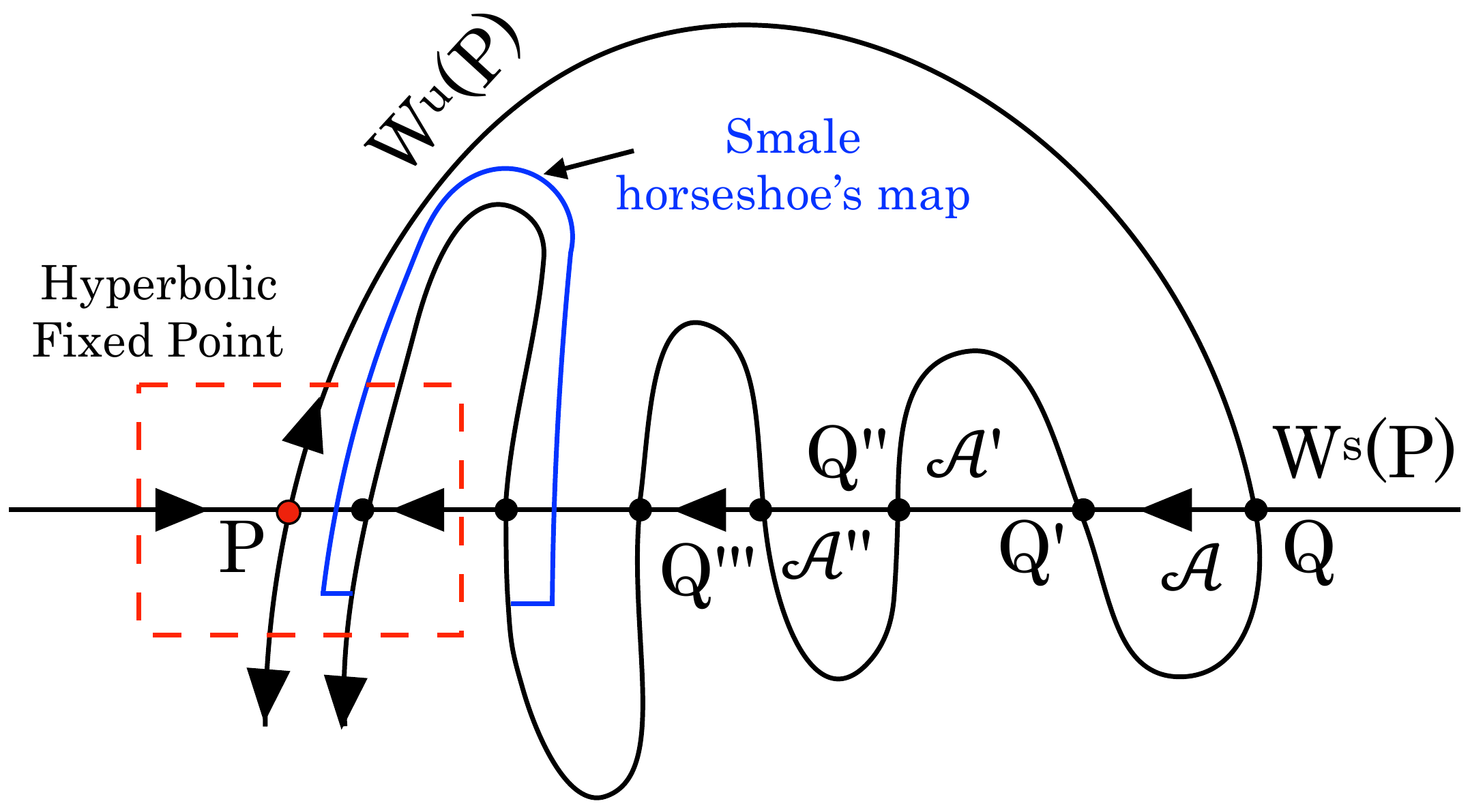}
\caption{Example of homoclinic tangle.}
\label{fig:Fig4}
\end{figure}

Let $\boldsymbol{\Phi}:\mathbb{R}^{2n}\to\mathbb{R}^{2n}$ be an Hamiltonian integrable dynamical system, which for the Liouville theorem is an area-preserving map in the phase space, possessing a hyperbolic fixed point $P$ and a homoclinic orbit $\mathcal{O}$. Such a Hamiltonian system is affected by dissipative perturbations like Eqs. (\ref{eq:Hmod}). In these hypothesis, the Melnikov method goes in search of \emph{homoclinic tangles} \cite{Guckenheimer2002,Wiggins1988,Bombelli1992}, see Fig. \ref{fig:Fig4}.
In such structures, once $W^s(P)$ and $W^u(P)$ intersect each other, they will continue to intersect infinitely in a discrete number of points, like $\left\{ Q,Q',Q'',Q''',\cdots,P\right\}$. They cannot touch the same point twice, otherwise they will be trapped in a cycle and will not reach the point $P$, and $P$ is not touched in a finite number of steps, since $P$ is a fixed point (has no image or pre-image of a point other than itself). Since $\boldsymbol{\Phi}$ is an area-preserving map, the areas formed by the intersection of $W^s(P)$ and $W^u(P)$  (i.e., $\mathcal{A},\mathcal{A}',\mathcal{A}'',\cdots$) are all equivalent. The \emph{Smale-Birkhoff theorem} claims that the dynamics produced by $W^s(P)$ and $W^u(P)$ in approaching the point $P$ gives rise to the \emph{Smale horseshoe's map}, which is a \emph{chaotic map} \cite{Guckenheimer2002,Wiggins1988}.

\begin{figure}[h!]
\includegraphics[scale=0.4]{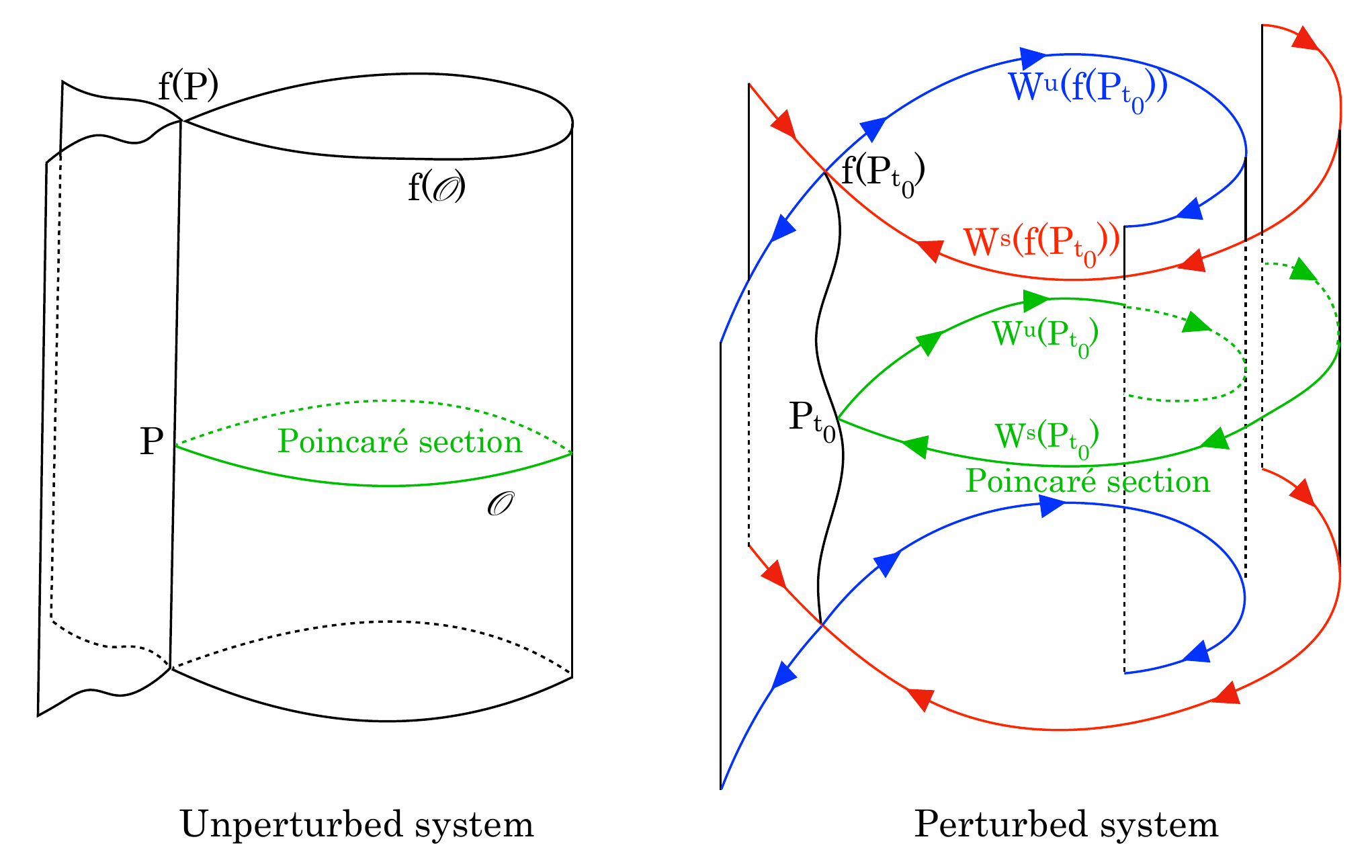}
\caption{Sketch of Poincar\'e sections for unperturbed (left cartoon) and perturbed (right cartoon) dynamical systems.}
\label{fig:Plot2}
\end{figure}

\begin{figure}[h!]
\includegraphics[scale=0.35]{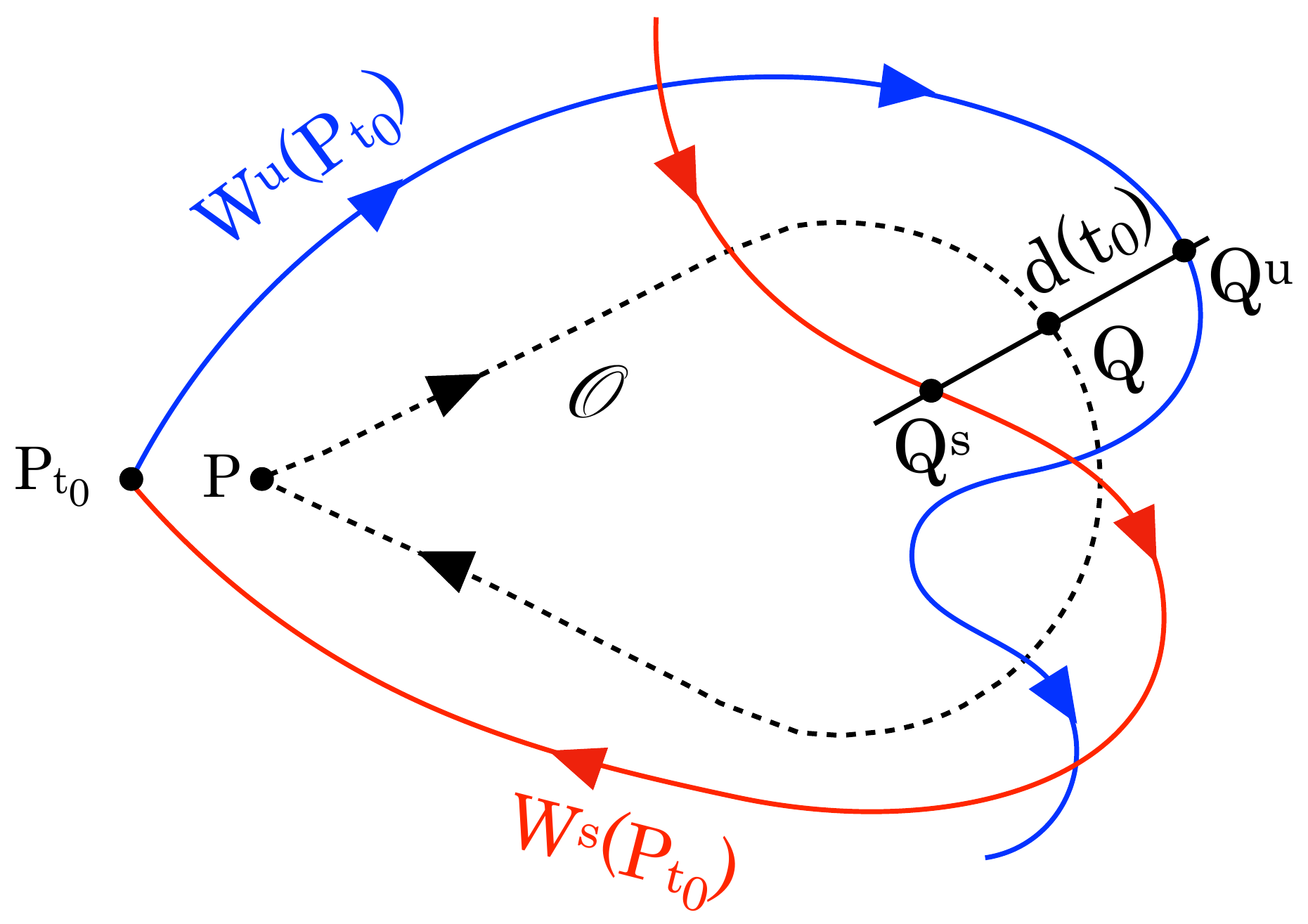}
\caption{Cartoon to explain the Melnikov integral.}
\label{fig:Fig5}
\end{figure}

In order to find the homoclinic tangles, we have to find a time $t_0$ such that $W^s(P_{t_0})$ and $W^u(P_{t_0})$ intersect transversally. To this end, we fix an arbitrary initial time $t_0$ or Poincar\'e section (where we follow the dynamics) corresponding to the hyperbolic fixed point $P_{t_0}$ (see Fig. \ref{fig:Plot2}). 
We consider a point $Q\in\mathcal{O}$, and then we define the distance from $W^s(P_{t_0})$ to $W^u(P_{t_0})$ along a transversal direction to $\mathcal{O}$ in $Q$, which intersects $W^s(P_{t_0})$ and $W^u(P_{t_0})$ respectively in $Q^s$ and $Q^u$, see Fig. \ref{fig:Fig5},
\begin{equation} 
d(t_0)\approx \epsilon \frac{\mathcal{M}(t_0)}{||\nabla \mathcal{H}(P_{t_0})||}+O(\epsilon^2),
\end{equation}
where $||\nabla \mathcal{H}(P_{t_0})||\neq0$. Here $\mathcal{M}(t_0)$ is the \emph{Melnikov integral} defined as \cite{Guckenheimer2002,Wiggins1988,Holmes1982V1,Holmes1982V2,Bombelli1992}
\begin{equation}\label{eq:MI}
\begin{aligned}
\mathcal{M}(t_0)&=\int_{-\infty}^{+\infty}\left\{\mathcal{H},\boldsymbol{f}\right\}\ dt\\
&=\int_{-\infty}^{+\infty}\sum_{\mu=1}^n\left(\frac{\partial\mathcal{H}}{\partial p_\mu}f_{2,\mu}+\frac{\partial\mathcal{H}}{\partial x^\mu}f_1^\mu\right)\ dt,
\end{aligned}
\end{equation}
where $\left\{\cdot,\cdot\right\}$ are the Poisson brackets, $\boldsymbol{f}$ are the perturbations of Eqs. (\ref{eq:Hmod}),
and the integral is taken along the unperturbed homoclinic orbit $\mathcal{O}$. Depending on the values assumed by the Melnikov integral, we have:
\begin{itemize}
\item if $\mathcal{M}(t_0)$ admits odd order zeros, there is chaos;  
\item if $\mathcal{M}(t_0)$ is bounded away from zero, there is no occurrence of chaos in the perturbed dynamics;
\item if $\mathcal{M}(t_0)$ is identically zero or admits even order zeros, the method cannot predict anything. 
\end{itemize}
 
\subsection{Application to the general relativistic PR effect perturbing the equatorial Kerr dynamics}
\label{sec:PRinMM}
We apply the Melnikov method to the general relativistic PR effect, where the perturbations are $\boldsymbol{f}=(0,\tilde{F}_\mu)$, see Sec. \ref{sec:PRinHF}. The Melnikov integral (\ref{eq:MI}) reads as
\begin{equation} \label{eq:PRMI}
\begin{aligned}
&\mathcal{M}\equiv\mathcal{M}(a,r_u,R_\star,\Omega_\star;t_0)\\
&=\int_{-\infty}^{+\infty}\left[\left(g^{tt}p_t+g^{t\varphi}p_\varphi\right)_{t=t_0}\tilde{F}_t(t-t_0)\right.\\
&\left.+\left(g^{\varphi\varphi}p_\varphi+g^{t\varphi}p_t\right)_{t=t_0}\tilde{F}_\varphi(t-t_0)\right.\\
&\left.+\left(g^{rr}p_r\right)_{t=t_0}\tilde{F}_r(t-t_0)\right]\ dt.
\end{aligned}
\end{equation}
This integral is evaluated along the homoclinic orbit $\mathcal{O}\in\mathcal{O}^{\rm hc}(r_u)$ at the time $t_0$. It is important to note that if there exists an intersection for some $t_0$, then there will be one for every $t_0$ \cite{Bombelli1992}.
Considering $p_r=\dot{r}g_{rr}$, $p_\varphi=L_z$, and $p_t=-E$ (see Eqs. (\ref{eq:r}), (\ref{eq:ampart}), and (\ref{eq:enpart}), respectively), the explicit expressions of $\hat{\mathcal{V}}^{\hat\mu}$ (see Eqs. (\ref{eq:FACT}) -- (\ref{eq:Vt})), and passing from the coordinate time $t$ to the coordinate radius $r$ integration, cf. Eqs. (\ref{eq:r}), we have
\begin{equation}
\mathcal{M}=2\int_{r_u}^{r_a}\Psi_1\Psi_2 dr,
\end{equation}
where
\begin{eqnarray}
\frac{\dot{t}}{\dot{r}}&=&\frac{r(r\rho E-2aL_z)}{\Delta\sqrt{R(r)}},\label{eq:JT}\\
\Psi_1&=&\frac{\gamma^2 (1+bN^\varphi)^2}{N^2\sqrt{R_{\rm rad}}}\frac{\mathbb{A}}{r_0\Delta_0}\frac{\dot{t}}{\dot{r}},\\
\Psi_2&=&(\rho_0r_0E-2aL_z)(1-\gamma\mathbb{A})\\
&&+\left[(r_0-2)L_z+2aE\right]\left(\cos\beta-\mathcal{B}\mathbb{A}\right)\notag\\
&&+\frac{\sqrt{R_0}\Delta_0}{r_0}\left(\sin\beta-\mathcal{A}\mathbb{A}\right).\notag
\end{eqnarray}
In order to simplify the notations we have defined
\begin{equation} \label{eq:AB}
\mathcal{A}=\frac{p_r}{\sqrt{g_{rr}}}\equiv\frac{1}{r}\sqrt{\frac{R}{\Delta}}\geq0,\quad \mathcal{B}=\frac{L_z}{\sqrt{g_{\varphi\varphi}}}\equiv\frac{L_z}{\sqrt{\rho}}>0.
\end{equation}
This implies that Eqs. (\ref{eq:velga}) -- (\ref{eq:AA}) reads respectively as
\begin{equation} \label{eq:AB}
\gamma=\sqrt{1+\mathcal{A}^2+\mathcal{B}^2},\quad 
\mathbb{A}=\gamma-\mathcal{A}\sin\beta-\mathcal{B}\cos\beta.
\end{equation}
The quantities with a subscript zero means that they are evaluated at the initial time $t_0$. Since the general relativistic PR effect dynamics does not depend explicitly on the time, we can set without loss of generality and for simplicity $t_0=0$. We will show that $\mathcal{M}$ has not a defined sign for all parameters ranging in their intervals. To achieve this goal, we will study the signs of each components of the integrating function by performing either analytical calculations or numerical simulations, where the functions are difficult to handle analytically. 

Resuming what has been discussed in the previous sections, we have that the set of parameters $\left\{a,r_u,R_\star,\Omega_\star,r_0,r,\beta\right\}$ range over the following intervals
\begin{equation} \label{eq:ranges}
\begin{aligned}
&a\in[0,1),\qquad  r_u\in[r_{\rm IBCO}(a),r_{\rm ISCO}(a)],\\
& R_\star\in(r_{\rm H}(a),\bar{R}_\star],\qquad \Omega_\star\in[\Omega_{\rm min},\Omega_{\rm max}],\\
&r_0\in(r_{\rm H}(a),\bar{R}], \quad  r\in[r_u,r_a(a,r_u)],\quad \beta\in[0,2\pi],
\end{aligned}
\end{equation}
where $\bar{R}_\star,\bar{R}$ are finite values. In addition, we know
\begin{equation}
\frac{\sqrt{3}}{3}\leq E<1,\quad \frac{2\sqrt{3}}{3}\leq L_z\leq4,\quad \gamma\geq1.
\end{equation}

We prove that $\Psi_1>0$, because  
\begin{equation}
\frac{\gamma^2 (1+bN^\varphi)^2}{N^2\sqrt{R_{\rm rad}}}>0, \quad \frac{\mathbb{A}}{r_0\Delta_0}>0, \quad 
\frac{\dot{t}}{\dot{r}}>0.
\end{equation}
The first term is composed by positive quantities. The second term is positive, because $\mathbb{A}>0$ (see Appendix \ref{app:AA} and Eqs. (\ref{eq:AB})). The term $\dot{t}/\dot{r}$ has a positive denominator, as well as the numerator (see the proof in Appendix \ref{app:J}). 

We prove that $\Psi_2$ has not a defined sign. Indeed, we have that $\rho_0r_0E-2aL_z$ is equal to the numerator of $\dot{t}/\dot{r}$, which is non-negative, $\sqrt{R_0}\Delta_0/r_0$ is non-negative and
\begin{equation}
(r_0-2)L_z+2aE>(1-2)\frac{2\sqrt{3}}{3}+\frac{2\sqrt{3}}{3}=0.
\end{equation} 
\begin{figure*}[th!]
\centering
\hbox{\includegraphics[scale=0.26]{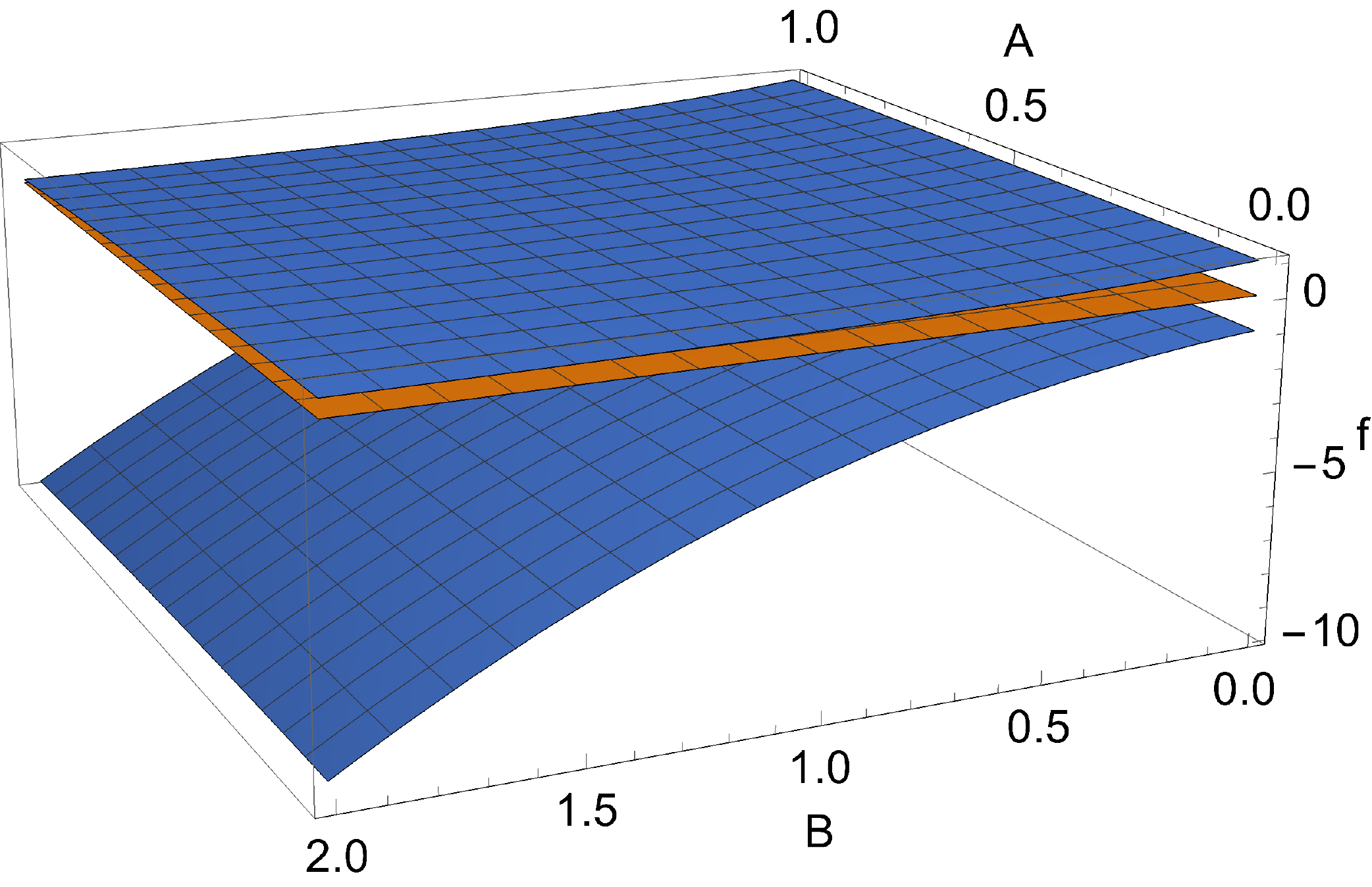}
\hspace{0.2cm}
\includegraphics[scale=0.26]{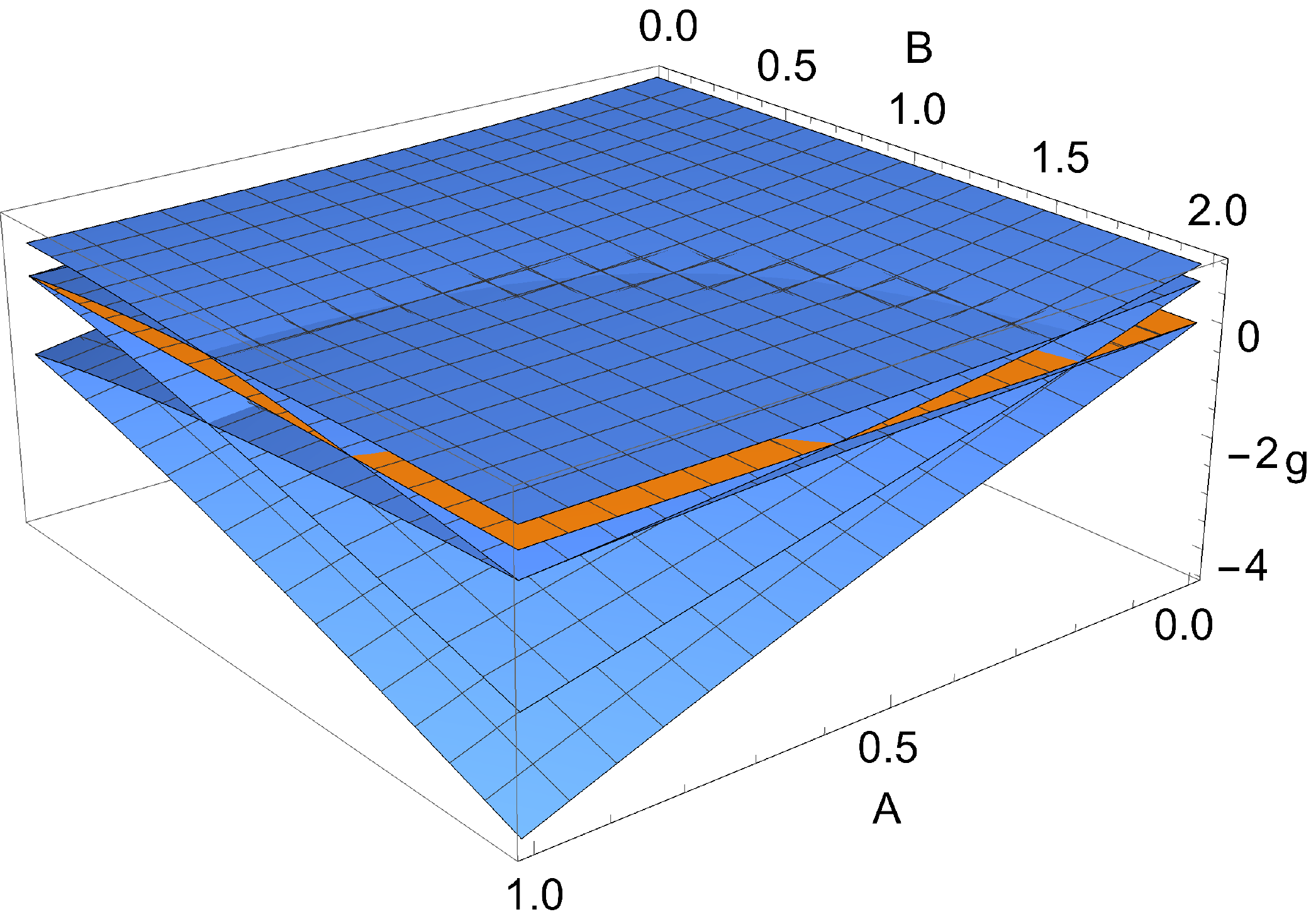}
\hspace{0.2cm}
\includegraphics[scale=0.26]{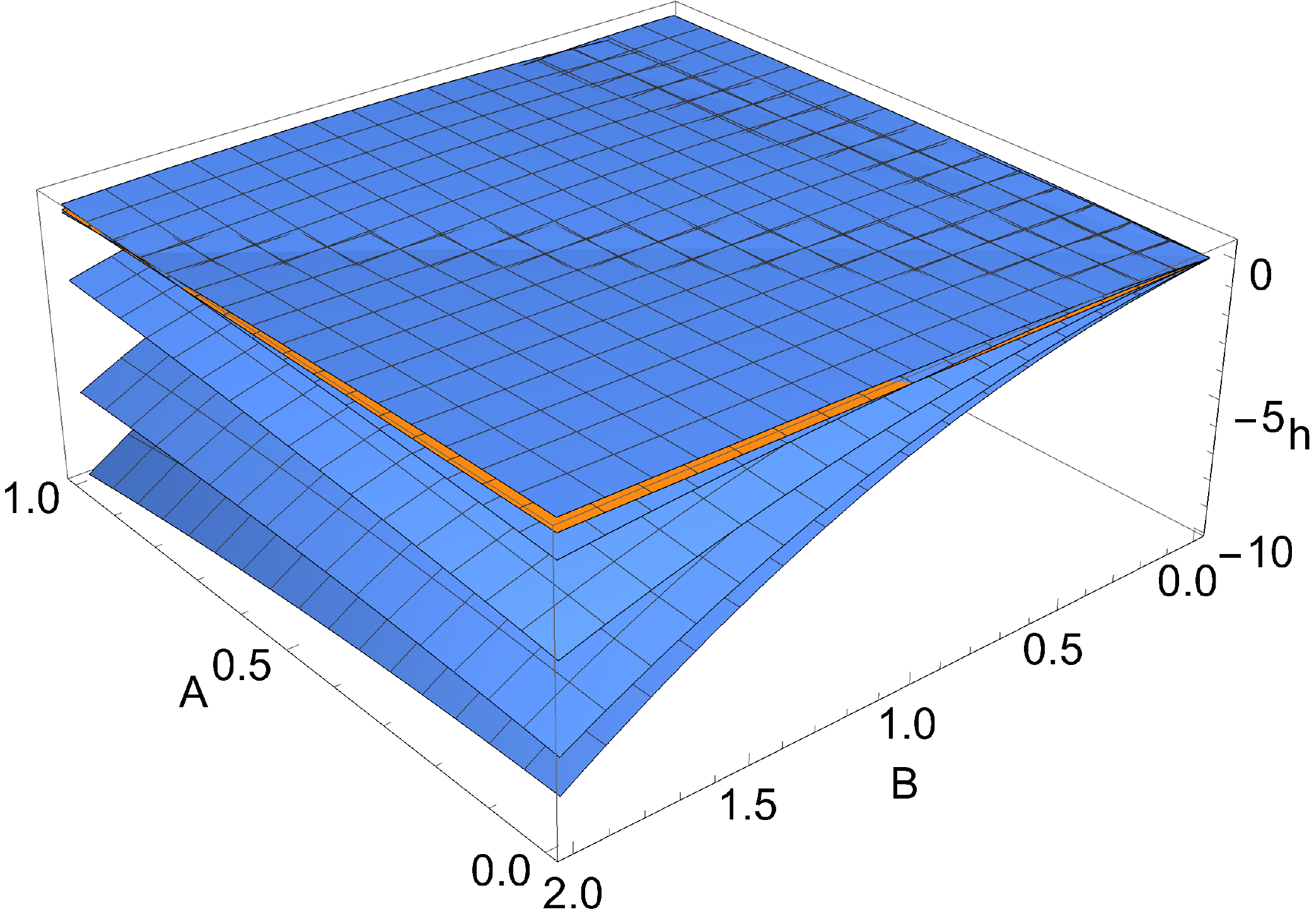}}
\caption{Ranges of the $f,g,h$ functions (blue surfaces). The orange surface corresponds to zero value of the functions.}
\label{fig:FigR}
\end{figure*}

The signs of $\Psi_2$ depends therefore only by (defining $f=\cos\beta-\mathcal{B}\mathbb{A}$, $g=\sin\beta-\mathcal{A}\mathbb{A}$, $h=1-\gamma\mathbb{A}$)
\begin{eqnarray}
f(\mathcal{A},\mathcal{B},\beta)&&=\cos\beta(1+\mathcal{B}^2)+\mathcal{B}(\mathcal{A}\sin\beta-\gamma),\\
g(\mathcal{A},\mathcal{B},\beta)&&=\sin\beta(1+\mathcal{A}^2)+\mathcal{A}(\mathcal{B}\cos\beta-\gamma),\\
h(\mathcal{A},\mathcal{B},\beta)&&=-(\mathcal{A}^2+\mathcal{B}^2)+\mathcal{A}\sqrt{1+\mathcal{A}^2+\mathcal{B}^2}.
\end{eqnarray}
After having found the ranges of $\mathcal{A},\mathcal{B}$ (see Appendix \ref{app:ABGAMMA}), we see that for $\beta\in[0,2\pi]$ the functions $f,g,h$ does not have a definite sign as can be seen in Fig. \ref{fig:FigR}

This result implies that the Melnikov integral may admit zero values. Due to the behaviors of the $f,g,h$ functions it is very difficult to analytically describe the set of parameters for which the Melnikov integral vanishes. Therefore, we resort to numerical simulations to investigate this issue.
We develop a code in \texttt{Mathematica 12.1.1.0}, which permits to numerically check whether there are values of $r_0$ such that vanish the Melnikov integral. We calculate also the derivative of the Melnikov integral with respect to $r_0$, proving that it is non-zero at the value of $r_0$ for which the Melnikov integral vanishes, thus assuring that the zero is simple and chaos occurs (see {\it Theorem 4.5.2} in Ref. \cite{Guckenheimer2002}, for more details) \footnote{We note that our dynamical system is autonomous with respect to the time $t$. Since the model is set in the equatorial plane all the functions depend only by $r(t)$. Therefore, we should find the value of $r_0$ such that $\mathcal{M}(r_0)=0$. In addition the condition
$$
\frac{\partial \mathcal{M}}{\partial t_0}\neq0,
$$
can be substituted by
$$
\frac{\partial \mathcal{M}}{\partial r_0}\left[\frac{\partial r(t)}{\partial t}\right]_{t=t_0}.
$$}.

\begin{figure}[ht!] 
\centering
\includegraphics[scale=0.3]{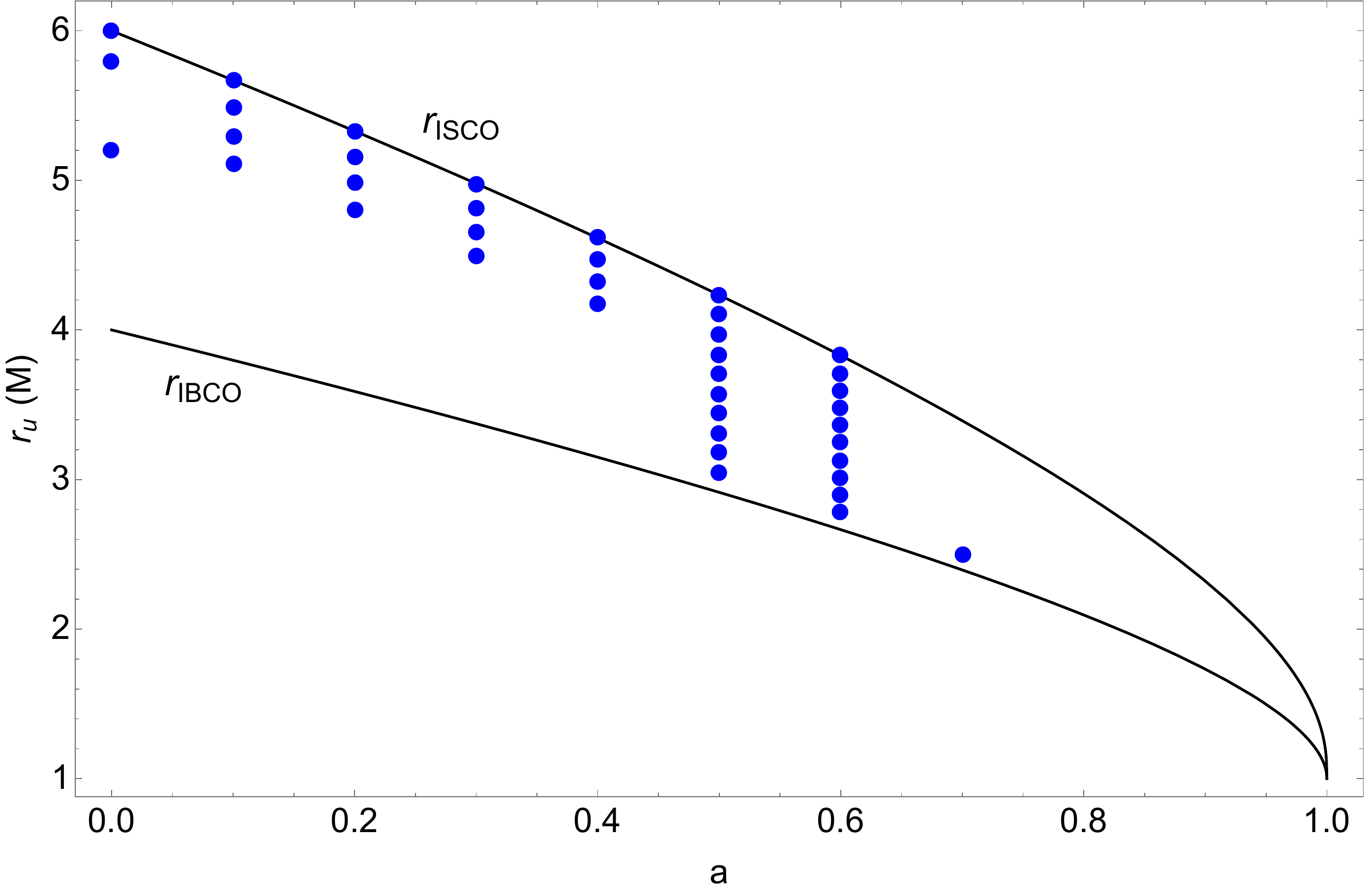}
\caption{Parameter space $(a,r_u)$ delimited by the curves $r_{\rm IBCO}$ and $r_{\rm ISCO}$ for the photon impact parameter value $b=3$. The blue dots are the values found from our numerical simulations in \texttt{Mathematica}, corresponding to chaotic dynamics.}
\label{fig:Fig6}
\end{figure}

\begin{figure}[ht!] 
\centering
\includegraphics[trim=0.3cm 1.5cm 1.5cm 0.3cm, scale=0.345]{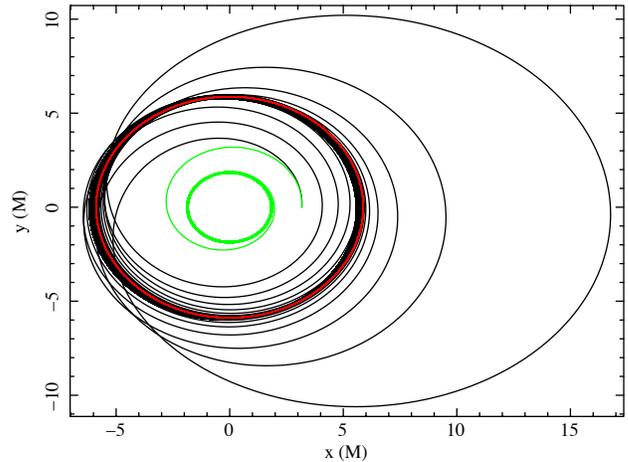}
\caption{Trajectories of two distinct test particles for $a=0.5$, $A=L/L_{\rm Edd}=0.1$, $b=3.1M$ starting both at $r_0=3.17M$ with angular velocity $\alpha_0=0.11$, but with different initial velocities, $\nu_0=0.67$ (black and chaotic orbit) and $\nu_0=0.60$ (green orbit). The red circle is the critical hypersurface for the black orbit located at $r_{\rm crit}=5.88M$, while the critical hypersurface for the green one is located at $r_{\rm crit}=1.88M$, very close to the event horizon $R_{\rm H}(0.5)=1.87M$.}
\label{fig:Fig7}
\end{figure}

In our numerical simulations we found chaotic behavior for certain parameter values. In Fig. \ref{fig:Fig6}, we display the region of the parameter space where chaos occurs.
We performed several numerical simulations, and we found that for photon impact parameter $b=3$ there is the occurrence of chaos for almost each spin value, while for $b=0,1,2$ chaos is not present. \emph{Therefore, we conclude that for radial radiation field, $b=0$, chaos does not manifest}. In addition, we checked that for values closer to $b=3$, chaos still reveals its presence. Once $(a^*,r_u^*)$, and the radius $r_0$ for having chaotic dynamics have been found, we calculate the initial conditions on the test particle trajectory by calculating first $\mathcal{A}^*=\mathcal{A}(a^*r_u^*,r_0)$ and $\mathcal{B}^*=\mathcal{B}(a^*r_u^*,r_0)$, see Eqs. (\ref{eq:AB}), and then we can calculate the test particle initial velocity conditions 
\begin{equation}
\nu_0=\sqrt{\frac{\mathcal{A}^*{}^2+\mathcal{B}^*{}^2}{1+\mathcal{A}^*{}^2+\mathcal{B}^*{}^2}},\qquad \alpha_0=\arctan\left(\frac{\mathcal{A}^*{}}{\mathcal{B}^*{}}\right),
\end{equation}
obtained by employing Eqs. (\ref{eq:prf}), (\ref{eq:velga}), and (\ref{eq:AB}). 
As an example we plot in Fig. \ref{fig:Fig7} a chaotic orbit together with a normal dynamics to highlight the different behaviours. 

\section{Conclusions}
\label{sec:end}
We have analysed the general relativistic PR effect in the equatorial plane of Kerr spacetime from a dynamical system point of view (see Sec. \ref{sec:PRmodel}). We have employed the Melnikov method to investigate whether it admits chaotic behaviors. The idea behind such investigation relies on the existence of some \emph{a-priori} indications of chaos, which are: $(i)$ non-integrability, $(ii)$ presence of a stable attractor (critical hypersurface), $(iii)$ strong analogy with a forced harmonic oscillator (Duffing equations), $(iv)$ sensitive dependence on the initial conditions (see Sec. \ref{sec:ICB}). 

The Melnikov method is based on the knowledge of unperturbed Hamiltonian Kerr metric, general relativistic PR dissipative perturbations (see Sec. \ref{sec:PRinHF}), and homoclinic orbits in the equatorial plane of Kerr spacetime, parametrized by the periastron $r_u$ (see Sec. \ref{sec:Levin}). The aim is for the existence of homoclinic tangles in the phase space, whose dynamics reproduce that of the Smale horseshoe's map, which is a chaotic map. This reduces to determining whether the Melnikov integral, see Eq. (\ref{eq:PRMI}), admits zeros in terms of its parameters $(a,r_u,b(R_\star,\Omega_\star))$ and initial condition $r_0$, see Sec. \ref{sec:MI}. We proved that the Melnikov integral admits simple zeros and therefore chaos is present in the dynamics of the general relativistic PR effect in the Kerr equatorial plane for low luminosities.

This result is relevant, because we discovered that the general relativistic PR effect can admit chaotic orbits for a suitable range of parameters provided by the Melnikov method. Although several numerical simulations of the PR trajectories have been performed in the literature (see Refs. \cite{Bini2009,Bini2011,DeFalcoTESI}, for further details), to our knowledge the existence of chaotic trajectories has never been reported in previous works on the PR effect. These configurations are useful for astrophysical purposes, because they can be exploited as a valuable tool for lighting up the compact object around which a test particle orbits, being thus a further source of information \cite{Levin1999}. Instead, for observational goals it could be better to avoid chaotic orbits for not compromising the detections of phenomena related to the general relativistic PR effect \cite{Cornish2001}. 

\section*{Acknowledgements}
V.D.F. thanks Gruppo Nazionale di Fisica Matematica of Istituto Nazionale di Alta Matematica for support. W.B. acknowledges support from Gruppo Nazionale per l'Analisi Matematica, la Probabilit\`a e le loro Applicazioni of Istituto Nazionale di Alta Matematica. V.D.F. and W.B. are grateful to Dr. Paolo Giulietti for fruitful discussions on the Melnikov method and on Chaos Theory.

\begin{appendix}
\section{signs and ranges of parameters}
\label{Appendix:SR}
This Appendix is devoted to prove the sign or range of some parameters, appearing in the terms $\Psi_1,\Psi_2$.
\subsection{Sign and range of $\mathbb{A}$}
\label{app:AA}
The expression of $\mathbb{A}$, see Eq. (\ref{eq:AA}), can be equivalently written in terms of Eqs. (\ref{eq:AB}) as
\begin{equation}
\label{eq:A}
\mathbb{A}=\sqrt{1+\mathcal{A}^2+\mathcal{B}^2}-\mathcal{A}\sin\beta-\mathcal{B}\cos\beta,
\end{equation}
where $\gamma=\sqrt{1+\mathcal{A}^2+\mathcal{B}^2}$, cf. Eqs. (\ref{eq:velga}). We claim that $\mathbb{A}>0$. If $\sin\beta,\ \cos\beta\leq0$ this derives immediately from \eqref{eq:A}. Assuming that $\sin\beta>0$ and $\cos\beta\leq 0$ we get 
\[
\mathbb{A}\geq \sqrt{1+\mathcal{A}^2+\mathcal{B}^2}-\mathcal{A}\sin\beta>\sqrt{1+\mathcal{A}^2}-\mathcal{A}>0\,
\]
and the same argument holds exchanging the role of $\cos\beta$ and $\sin\beta$. Finally, if $\cos\beta,\ \sin\beta>0$, we obtain
\begin{equation}
1+\mathcal{A}^2+\mathcal{B}^2> (\mathcal{A}\sin\beta+\mathcal{B}\cos\beta)^2,
\end{equation}
from which we obtain
\begin{equation}
1+(\mathcal{A}\cos\beta-\mathcal{B}\sin\beta)^2>0.
\end{equation}

\subsection{Sign of $\dot{t}/\dot{r}$}
\label{app:J}
The term $\dot{t}/\dot{r}$ is the Jacobian of coordinate transformation, cf. Eq. (\ref{eq:JT}), therefore it must be non-zero. Since the denominator is always positive, we focus only on the sign of the numerator, which is estimated through
\begin{equation}
r\rho E-2aL_z\geq r\rho \frac{\sqrt{3}}{3}-\frac{4\sqrt{3}}{3}>\frac{4\sqrt{3}}{3}-\frac{4\sqrt{3}}{3}=0.
\end{equation}

\subsection{Ranges of $\mathcal{A}$ and $\mathcal{B}$}
\label{app:ABGAMMA}
The analytical expression of $\mathcal{A}$ is, see Eqs. (\ref{eq:FR}) -- (\ref{eq:AB}),
\begin{equation}
\mathcal{A}=\sqrt{\frac{(1-E^2)(r-r_u)^2(r_a-r)}{r\Delta}}.
\end{equation}
where the numerator has a maximum at $r_m=(r_u+2r_a)/3$, while the denominator is a monotone increasing function for $r\in[r_u,r_a]$. We numerically checked that $\mathcal{A}$ attains its maximum for $r_u= R_{\rm IBCO}(a)$. It becomes a constant function independent from $a$, such that $\mathcal{A}(a,R_{\rm IBCO}(a))\approx0.7$, showing thus that $\mathcal{A}<1$. 

The range of $\mathcal{B}$ can be obtained through
\begin{equation} 
0\leq\mathcal{B}\leq\frac{L_z}{r_u}<2.
\end{equation}

\end{appendix}
\bibliography{references}

\begin{thebibliography}{45}%
\makeatletter
\providecommand \@ifxundefined [1]{%
 \@ifx{#1\undefined}
}%
\providecommand \@ifnum [1]{%
 \ifnum #1\expandafter \@firstoftwo
 \else \expandafter \@secondoftwo
 \fi
}%
\providecommand \@ifx [1]{%
 \ifx #1\expandafter \@firstoftwo
 \else \expandafter \@secondoftwo
 \fi
}%
\providecommand \natexlab [1]{#1}%
\providecommand \enquote  [1]{``#1''}%
\providecommand \bibnamefont  [1]{#1}%
\providecommand \bibfnamefont [1]{#1}%
\providecommand \citenamefont [1]{#1}%
\providecommand \href@noop [0]{\@secondoftwo}%
\providecommand \href [0]{\begingroup \@sanitize@url \@href}%
\providecommand \@href[1]{\@@startlink{#1}\@@href}%
\providecommand \@@href[1]{\endgroup#1\@@endlink}%
\providecommand \@sanitize@url [0]{\catcode `\\12\catcode `\$12\catcode
  `\&12\catcode `\#12\catcode `\^12\catcode `\_12\catcode `\%12\relax}%
\providecommand \@@startlink[1]{}%
\providecommand \@@endlink[0]{}%
\providecommand \url  [0]{\begingroup\@sanitize@url \@url }%
\providecommand \@url [1]{\endgroup\@href {#1}{\urlprefix }}%
\providecommand \urlprefix  [0]{URL }%
\providecommand \Eprint [0]{\href }%
\providecommand \doibase [0]{http://dx.doi.org/}%
\providecommand \selectlanguage [0]{\@gobble}%
\providecommand \bibinfo  [0]{\@secondoftwo}%
\providecommand \bibfield  [0]{\@secondoftwo}%
\providecommand \translation [1]{[#1]}%
\providecommand \BibitemOpen [0]{}%
\providecommand \bibitemStop [0]{}%
\providecommand \bibitemNoStop [0]{.\EOS\space}%
\providecommand \EOS [0]{\spacefactor3000\relax}%
\providecommand \BibitemShut  [1]{\csname bibitem#1\endcsname}%
\let\auto@bib@innerbib\@empty
\bibitem [{\citenamefont {Wiggins}(1988)}]{Wiggins1988}%
  \BibitemOpen
  \bibfield  {author} {\bibinfo {author} {\bibfnamefont {S.}~\bibnamefont
  {Wiggins}},\ }\href {https://books.google.it/books?id=fO5QAAAAMAAJ} {\emph
  {\bibinfo {title} {Global Bifurcations and Chaos: Analytical Methods}}},\
  \bibinfo {series} {Applied Mathematical Sciences Series}\ No.\ \bibinfo
  {number} {v. 73}\ (\bibinfo  {publisher} {Springer-Verlag},\ \bibinfo {year}
  {1988})\BibitemShut {NoStop}%
\bibitem [{\citenamefont {Ott}(2002)}]{Ott2002}%
  \BibitemOpen
  \bibfield  {author} {\bibinfo {author} {\bibfnamefont {E.}~\bibnamefont
  {Ott}},\ }\href {https://books.google.it/books?id=nOLx--zzHSgC} {\emph
  {\bibinfo {title} {Chaos in Dynamical Systems}}}\ (\bibinfo  {publisher}
  {Cambridge University Press},\ \bibinfo {year} {2002})\BibitemShut {NoStop}%
\bibitem [{\citenamefont {Devaney}(2018)}]{Devaney2018}%
  \BibitemOpen
  \bibfield  {author} {\bibinfo {author} {\bibfnamefont {R.}~\bibnamefont
  {Devaney}},\ }\href@noop {} {\emph {\bibinfo {title} {An introduction to
  chaotic dynamical systems}}}\ (\bibinfo  {publisher} {CRC Press},\ \bibinfo
  {year} {2018})\BibitemShut {NoStop}%
\bibitem [{\citenamefont {Hobill}\ \emph {et~al.}(1994)\citenamefont {Hobill},
  \citenamefont {Burd},\ and\ \citenamefont {Coley}}]{Hobill1994}%
  \BibitemOpen
  \bibfield  {author} {\bibinfo {author} {\bibfnamefont {D.}~\bibnamefont
  {Hobill}}, \bibinfo {author} {\bibfnamefont {A.}~\bibnamefont {Burd}}, \ and\
  \bibinfo {author} {\bibfnamefont {A.}~\bibnamefont {Coley}},\ }\href
  {https://books.google.it/books?id=LUgsAAAAYAAJ} {\emph {\bibinfo {title}
  {Deterministic Chaos in General Relativity}}},\ Nato Science Series B:\
  (\bibinfo  {publisher} {Springer US},\ \bibinfo {year} {1994})\BibitemShut
  {NoStop}%
\bibitem [{\citenamefont {Arnol'd}\ and\ \citenamefont
  {Avez}(1989)}]{Arnold1989}%
  \BibitemOpen
  \bibfield  {author} {\bibinfo {author} {\bibfnamefont {V.}~\bibnamefont
  {Arnol'd}}\ and\ \bibinfo {author} {\bibfnamefont {A.}~\bibnamefont {Avez}},\
  }\href {https://books.google.it/books?id=9TxhQgAACAAJ} {\emph {\bibinfo
  {title} {Ergodic Problems of Classical Mechanics}}},\ Advanced book classics\
  (\bibinfo  {publisher} {Addison-Wesley},\ \bibinfo {year} {1989})\BibitemShut
  {NoStop}%
\bibitem [{\citenamefont {{Contopoulos}}(1990)}]{Contopoulos1990}%
  \BibitemOpen
  \bibfield  {author} {\bibinfo {author} {\bibfnamefont {G.}~\bibnamefont
  {{Contopoulos}}},\ }\href {\doibase 10.1098/rspa.1990.0126} {\bibfield
  {journal} {\bibinfo  {journal} {Proceedings of the Royal Society of London
  Series A}\ }\textbf {\bibinfo {volume} {431}},\ \bibinfo {pages} {183}
  (\bibinfo {year} {1990})}\BibitemShut {NoStop}%
\bibitem [{\citenamefont {{Contopoulos}}(1991)}]{Contopoulos1991}%
  \BibitemOpen
  \bibfield  {author} {\bibinfo {author} {\bibfnamefont {G.}~\bibnamefont
  {{Contopoulos}}},\ }\href {\doibase 10.1098/rspa.1991.0160} {\bibfield
  {journal} {\bibinfo  {journal} {Proceedings of the Royal Society of London
  Series A}\ }\textbf {\bibinfo {volume} {435}},\ \bibinfo {pages} {551}
  (\bibinfo {year} {1991})}\BibitemShut {NoStop}%
\bibitem [{\citenamefont {{Wanex}}(2002)}]{Wanex2002}%
  \BibitemOpen
  \bibfield  {author} {\bibinfo {author} {\bibfnamefont {L.~F.}\ \bibnamefont
  {{Wanex}}},\ }\emph {\bibinfo {title} {{Chaotic amplification in the
  relativistic restricted three-body problem}}},\ \href@noop {} {Ph.D.
  thesis},\ \bibinfo  {school} {UNIVERSITY OF NEVADA, RENO} (\bibinfo {year}
  {2002})\BibitemShut {NoStop}%
\bibitem [{\citenamefont {{Bombelli}}\ and\ \citenamefont
  {{Calzetta}}(1992)}]{Bombelli1992}%
  \BibitemOpen
  \bibfield  {author} {\bibinfo {author} {\bibfnamefont {L.}~\bibnamefont
  {{Bombelli}}}\ and\ \bibinfo {author} {\bibfnamefont {E.}~\bibnamefont
  {{Calzetta}}},\ }\href {\doibase 10.1088/0264-9381/9/12/004} {\bibfield
  {journal} {\bibinfo  {journal} {Classical and Quantum Gravity}\ }\textbf
  {\bibinfo {volume} {9}},\ \bibinfo {pages} {2573} (\bibinfo {year}
  {1992})}\BibitemShut {NoStop}%
\bibitem [{\citenamefont {{Suzuki}}\ and\ \citenamefont
  {{Maeda}}(1997)}]{Suzuki1997}%
  \BibitemOpen
  \bibfield  {author} {\bibinfo {author} {\bibfnamefont {S.}~\bibnamefont
  {{Suzuki}}}\ and\ \bibinfo {author} {\bibfnamefont {K.-I.}\ \bibnamefont
  {{Maeda}}},\ }\href {\doibase 10.1103/PhysRevD.55.4848} {\bibfield  {journal}
  {\bibinfo  {journal} {PRD}\ }\textbf {\bibinfo {volume} {55}},\ \bibinfo
  {pages} {4848} (\bibinfo {year} {1997})},\ \Eprint
  {http://arxiv.org/abs/gr-qc/9604020} {arXiv:gr-qc/9604020 [gr-qc]}
  \BibitemShut {NoStop}%
\bibitem [{\citenamefont {{Lukes-Gerakopoulos}}(2018)}]{Lukes2018}%
  \BibitemOpen
  \bibfield  {author} {\bibinfo {author} {\bibfnamefont {G.}~\bibnamefont
  {{Lukes-Gerakopoulos}}},\ }in\ \href {\doibase 10.1142/9789813226609_0209}
  {\emph {\bibinfo {booktitle} {Fifteenth Marcel Grossmann Meeting - MG15}}}\
  (\bibinfo {year} {2018})\ pp.\ \bibinfo {pages} {1960--1965},\ \Eprint
  {http://arxiv.org/abs/1606.09430} {arXiv:1606.09430 [gr-qc]} \BibitemShut
  {NoStop}%
\bibitem [{\citenamefont {{Cornish}}(2001)}]{Cornish2001}%
  \BibitemOpen
  \bibfield  {author} {\bibinfo {author} {\bibfnamefont {N.~J.}\ \bibnamefont
  {{Cornish}}},\ }\href {\doibase 10.1103/PhysRevD.64.084011} {\bibfield
  {journal} {\bibinfo  {journal} {\prd}\ }\textbf {\bibinfo {volume} {64}},\
  \bibinfo {pages} {084011} (\bibinfo {year} {2001})},\ \Eprint
  {http://arxiv.org/abs/gr-qc/0106062} {arXiv:gr-qc/0106062 [gr-qc]}
  \BibitemShut {NoStop}%
\bibitem [{\citenamefont {{Cornish}}\ and\ \citenamefont
  {{Levin}}(2002)}]{Cornish2002}%
  \BibitemOpen
  \bibfield  {author} {\bibinfo {author} {\bibfnamefont {N.~J.}\ \bibnamefont
  {{Cornish}}}\ and\ \bibinfo {author} {\bibfnamefont {J.}~\bibnamefont
  {{Levin}}},\ }\href {\doibase 10.1103/PhysRevLett.89.179001} {\bibfield
  {journal} {\bibinfo  {journal} {\prl}\ }\textbf {\bibinfo {volume} {89}},\
  \bibinfo {eid} {179001} (\bibinfo {year} {2002})},\ \Eprint
  {http://arxiv.org/abs/gr-qc/0207020} {arXiv:gr-qc/0207020 [gr-qc]}
  \BibitemShut {NoStop}%
\bibitem [{\citenamefont {{Cornish}}\ and\ \citenamefont
  {{Levin}}(2003)}]{Cornish2003}%
  \BibitemOpen
  \bibfield  {author} {\bibinfo {author} {\bibfnamefont {N.~J.}\ \bibnamefont
  {{Cornish}}}\ and\ \bibinfo {author} {\bibfnamefont {J.}~\bibnamefont
  {{Levin}}},\ }\href {\doibase 10.1103/PhysRevD.68.024004} {\bibfield
  {journal} {\bibinfo  {journal} {\prd}\ }\textbf {\bibinfo {volume} {68}},\
  \bibinfo {eid} {024004} (\bibinfo {year} {2003})}\BibitemShut {NoStop}%
\bibitem [{\citenamefont {{Belinskij}}\ \emph
  {et~al.}(1970{\natexlab{a}})\citenamefont {{Belinskij}}, \citenamefont
  {{Lifshits}},\ and\ \citenamefont {{Khalatnikov}}}]{Belinskij1970a}%
  \BibitemOpen
  \bibfield  {author} {\bibinfo {author} {\bibfnamefont {V.~A.}\ \bibnamefont
  {{Belinskij}}}, \bibinfo {author} {\bibfnamefont {E.~M.}\ \bibnamefont
  {{Lifshits}}}, \ and\ \bibinfo {author} {\bibfnamefont {I.~M.}\ \bibnamefont
  {{Khalatnikov}}},\ }\href@noop {} {\bibfield  {journal} {\bibinfo  {journal}
  {Uspekhi Fizicheskikh Nauk}\ }\textbf {\bibinfo {volume} {102}},\ \bibinfo
  {pages} {463} (\bibinfo {year} {1970}{\natexlab{a}})}\BibitemShut {NoStop}%
\bibitem [{\citenamefont {{Belinskij}}\ \emph
  {et~al.}(1970{\natexlab{b}})\citenamefont {{Belinskij}}, \citenamefont
  {{Khalatnikov}},\ and\ \citenamefont {{Lifshits}}}]{Belinskij1970b}%
  \BibitemOpen
  \bibfield  {author} {\bibinfo {author} {\bibfnamefont {V.~A.}\ \bibnamefont
  {{Belinskij}}}, \bibinfo {author} {\bibfnamefont {I.~M.}\ \bibnamefont
  {{Khalatnikov}}}, \ and\ \bibinfo {author} {\bibfnamefont {E.~M.}\
  \bibnamefont {{Lifshits}}},\ }\href {\doibase 10.1080/00018737000101171}
  {\bibfield  {journal} {\bibinfo  {journal} {Advances in Physics}\ }\textbf
  {\bibinfo {volume} {19}},\ \bibinfo {pages} {525} (\bibinfo {year}
  {1970}{\natexlab{b}})}\BibitemShut {NoStop}%
\bibitem [{\citenamefont {{Barrow}}\ and\ \citenamefont
  {{Sirousse-Zia}}(1989)}]{Barrow1989}%
  \BibitemOpen
  \bibfield  {author} {\bibinfo {author} {\bibfnamefont {J.~D.}\ \bibnamefont
  {{Barrow}}}\ and\ \bibinfo {author} {\bibfnamefont {H.}~\bibnamefont
  {{Sirousse-Zia}}},\ }\href {\doibase 10.1103/PhysRevD.39.2187} {\bibfield
  {journal} {\bibinfo  {journal} {PRD}\ }\textbf {\bibinfo {volume} {39}},\
  \bibinfo {pages} {2187} (\bibinfo {year} {1989})}\BibitemShut {NoStop}%
\bibitem [{\citenamefont {{Burd}}\ \emph {et~al.}(1991)\citenamefont {{Burd}},
  \citenamefont {{Buric}},\ and\ \citenamefont {{Tavakol}}}]{Burd1991}%
  \BibitemOpen
  \bibfield  {author} {\bibinfo {author} {\bibfnamefont {A.~B.}\ \bibnamefont
  {{Burd}}}, \bibinfo {author} {\bibfnamefont {N.}~\bibnamefont {{Buric}}}, \
  and\ \bibinfo {author} {\bibfnamefont {R.~K.}\ \bibnamefont {{Tavakol}}},\
  }\href {\doibase 10.1088/0264-9381/8/1/014} {\bibfield  {journal} {\bibinfo
  {journal} {Classical and Quantum Gravity}\ }\textbf {\bibinfo {volume} {8}},\
  \bibinfo {pages} {123} (\bibinfo {year} {1991})}\BibitemShut {NoStop}%
\bibitem [{\citenamefont {{Contopoulos}}\ \emph {et~al.}(1999)\citenamefont
  {{Contopoulos}}, \citenamefont {{Voglis}},\ and\ \citenamefont
  {{Efthymiopoulos}}}]{Contopoulos1999}%
  \BibitemOpen
  \bibfield  {author} {\bibinfo {author} {\bibfnamefont {G.}~\bibnamefont
  {{Contopoulos}}}, \bibinfo {author} {\bibfnamefont {N.}~\bibnamefont
  {{Voglis}}}, \ and\ \bibinfo {author} {\bibfnamefont {C.}~\bibnamefont
  {{Efthymiopoulos}}},\ }\href {\doibase 10.1023/A:1008376523356} {\bibfield
  {journal} {\bibinfo  {journal} {Celestial Mechanics and Dynamical Astronomy}\
  }\textbf {\bibinfo {volume} {73}},\ \bibinfo {pages} {1} (\bibinfo {year}
  {1999})}\BibitemShut {NoStop}%
\bibitem [{\citenamefont {{Calzetta}}\ and\ \citenamefont {{El
  Hasi}}(1993)}]{Calzetta1993}%
  \BibitemOpen
  \bibfield  {author} {\bibinfo {author} {\bibfnamefont {E.}~\bibnamefont
  {{Calzetta}}}\ and\ \bibinfo {author} {\bibfnamefont {C.}~\bibnamefont {{El
  Hasi}}},\ }\href {\doibase 10.1088/0264-9381/10/9/022} {\bibfield  {journal}
  {\bibinfo  {journal} {Classical and Quantum Gravity}\ }\textbf {\bibinfo
  {volume} {10}},\ \bibinfo {pages} {1825} (\bibinfo {year} {1993})},\ \Eprint
  {http://arxiv.org/abs/gr-qc/9211027} {arXiv:gr-qc/9211027 [gr-qc]}
  \BibitemShut {NoStop}%
\bibitem [{\citenamefont {{Aydiner}}(2016)}]{Aydiner2016}%
  \BibitemOpen
  \bibfield  {author} {\bibinfo {author} {\bibfnamefont {E.}~\bibnamefont
  {{Aydiner}}},\ }\href@noop {} {\bibfield  {journal} {\bibinfo  {journal}
  {arXiv e-prints}\ ,\ \bibinfo {eid} {arXiv:1610.07338}} (\bibinfo {year}
  {2016})},\ \Eprint {http://arxiv.org/abs/1610.07338} {arXiv:1610.07338
  [gr-qc]} \BibitemShut {NoStop}%
\bibitem [{\citenamefont {{Poynting}}(1903)}]{Poynting1903}%
  \BibitemOpen
  \bibfield  {author} {\bibinfo {author} {\bibfnamefont {J.~H.}\ \bibnamefont
  {{Poynting}}},\ }\href@noop {} {\bibfield  {journal} {\bibinfo  {journal}
  {Monthly Notices of the Royal Astronomical Society}\ }\textbf {\bibinfo
  {volume} {64}},\ \bibinfo {pages} {1} (\bibinfo {year} {1903})}\BibitemShut
  {NoStop}%
\bibitem [{\citenamefont {{Robertson}}(1937)}]{Robertson1937}%
  \BibitemOpen
  \bibfield  {author} {\bibinfo {author} {\bibfnamefont {H.~P.}\ \bibnamefont
  {{Robertson}}},\ }\href {\doibase 10.1093/mnras/97.6.423} {\bibfield
  {journal} {\bibinfo  {journal} {Monthly Notices of the Royal Astronomical
  Society}\ }\textbf {\bibinfo {volume} {97}},\ \bibinfo {pages} {423}
  (\bibinfo {year} {1937})}\BibitemShut {NoStop}%
\bibitem [{\citenamefont {{Bini}}\ \emph {et~al.}(2009)\citenamefont {{Bini}},
  \citenamefont {{Jantzen}},\ and\ \citenamefont {{Stella}}}]{Bini2009}%
  \BibitemOpen
  \bibfield  {author} {\bibinfo {author} {\bibfnamefont {D.}~\bibnamefont
  {{Bini}}}, \bibinfo {author} {\bibfnamefont {R.~T.}\ \bibnamefont
  {{Jantzen}}}, \ and\ \bibinfo {author} {\bibfnamefont {L.}~\bibnamefont
  {{Stella}}},\ }\href {\doibase 10.1088/0264-9381/26/5/055009} {\bibfield
  {journal} {\bibinfo  {journal} {Classical and Quantum Gravity}\ }\textbf
  {\bibinfo {volume} {26}},\ \bibinfo {eid} {055009} (\bibinfo {year}
  {2009})},\ \Eprint {http://arxiv.org/abs/0808.1083} {arXiv:0808.1083 [gr-qc]}
  \BibitemShut {NoStop}%
\bibitem [{\citenamefont {{Bini}}\ \emph {et~al.}(2011)\citenamefont {{Bini}},
  \citenamefont {{Geralico}}, \citenamefont {{Jantzen}}, \citenamefont
  {{Semer{\'a}k}},\ and\ \citenamefont {{Stella}}}]{Bini2011}%
  \BibitemOpen
  \bibfield  {author} {\bibinfo {author} {\bibfnamefont {D.}~\bibnamefont
  {{Bini}}}, \bibinfo {author} {\bibfnamefont {A.}~\bibnamefont {{Geralico}}},
  \bibinfo {author} {\bibfnamefont {R.~T.}\ \bibnamefont {{Jantzen}}}, \bibinfo
  {author} {\bibfnamefont {O.}~\bibnamefont {{Semer{\'a}k}}}, \ and\ \bibinfo
  {author} {\bibfnamefont {L.}~\bibnamefont {{Stella}}},\ }\href {\doibase
  10.1088/0264-9381/28/3/035008} {\bibfield  {journal} {\bibinfo  {journal}
  {Classical and Quantum Gravity}\ }\textbf {\bibinfo {volume} {28}},\ \bibinfo
  {eid} {035008} (\bibinfo {year} {2011})},\ \Eprint
  {http://arxiv.org/abs/1408.4945} {arXiv:1408.4945 [gr-qc]} \BibitemShut
  {NoStop}%
\bibitem [{\citenamefont {{Bini}}\ \emph {et~al.}(2015)\citenamefont {{Bini}},
  \citenamefont {{Geralico}},\ and\ \citenamefont {{Passamonti}}}]{Bini2015}%
  \BibitemOpen
  \bibfield  {author} {\bibinfo {author} {\bibfnamefont {D.}~\bibnamefont
  {{Bini}}}, \bibinfo {author} {\bibfnamefont {A.}~\bibnamefont {{Geralico}}},
  \ and\ \bibinfo {author} {\bibfnamefont {A.}~\bibnamefont {{Passamonti}}},\
  }\href {\doibase 10.1093/mnras/stu2082} {\bibfield  {journal} {\bibinfo
  {journal} {MNRAS}\ }\textbf {\bibinfo {volume} {446}},\ \bibinfo {pages} {65}
  (\bibinfo {year} {2015})},\ \Eprint {http://arxiv.org/abs/1410.3099}
  {arXiv:1410.3099 [astro-ph.HE]} \BibitemShut {NoStop}%
\bibitem [{\citenamefont {{De Falco}}\ \emph {et~al.}(2019)\citenamefont {{De
  Falco}}, \citenamefont {{Bakala}}, \citenamefont {{Battista}}, \citenamefont
  {{Lan{\v c}ov{\'a}}}, \citenamefont {{Falanga}},\ and\ \citenamefont
  {{Stella}}}]{DeFalco20183D}%
  \BibitemOpen
  \bibfield  {author} {\bibinfo {author} {\bibfnamefont {V.}~\bibnamefont {{De
  Falco}}}, \bibinfo {author} {\bibfnamefont {P.}~\bibnamefont {{Bakala}}},
  \bibinfo {author} {\bibfnamefont {E.}~\bibnamefont {{Battista}}}, \bibinfo
  {author} {\bibfnamefont {D.}~\bibnamefont {{Lan{\v c}ov{\'a}}}}, \bibinfo
  {author} {\bibfnamefont {M.}~\bibnamefont {{Falanga}}}, \ and\ \bibinfo
  {author} {\bibfnamefont {L.}~\bibnamefont {{Stella}}},\ }\href {\doibase
  10.1103/PhysRevD.99.023014} {\bibfield  {journal} {\bibinfo  {journal}
  {\prd}\ }\textbf {\bibinfo {volume} {99}},\ \bibinfo {eid} {023014} (\bibinfo
  {year} {2019})}\BibitemShut {NoStop}%
\bibitem [{\citenamefont {{Bakala}}\ \emph {et~al.}(2019)\citenamefont
  {{Bakala}}, \citenamefont {{De Falco}}, \citenamefont {{Battista}},
  \citenamefont {{Goluchov{\'a}}}, \citenamefont {{Lan{\v{c}}ov{\'a}}},
  \citenamefont {{Falanga}},\ and\ \citenamefont {{Stella}}}]{Bakala2019}%
  \BibitemOpen
  \bibfield  {author} {\bibinfo {author} {\bibfnamefont {P.}~\bibnamefont
  {{Bakala}}}, \bibinfo {author} {\bibfnamefont {V.}~\bibnamefont {{De
  Falco}}}, \bibinfo {author} {\bibfnamefont {E.}~\bibnamefont {{Battista}}},
  \bibinfo {author} {\bibfnamefont {K.}~\bibnamefont {{Goluchov{\'a}}}},
  \bibinfo {author} {\bibfnamefont {D.}~\bibnamefont {{Lan{\v{c}}ov{\'a}}}},
  \bibinfo {author} {\bibfnamefont {M.}~\bibnamefont {{Falanga}}}, \ and\
  \bibinfo {author} {\bibfnamefont {L.}~\bibnamefont {{Stella}}},\ }\href
  {\doibase 10.1103/PhysRevD.100.104053} {\bibfield  {journal} {\bibinfo
  {journal} {PRD}\ }\textbf {\bibinfo {volume} {100}},\ \bibinfo {eid} {104053}
  (\bibinfo {year} {2019})}\BibitemShut {NoStop}%
\bibitem [{\citenamefont {{Wielgus}}(2019)}]{Wielgus2019}%
  \BibitemOpen
  \bibfield  {author} {\bibinfo {author} {\bibfnamefont {M.}~\bibnamefont
  {{Wielgus}}},\ }\href {\doibase 10.1093/mnras/stz2079} {\bibfield  {journal}
  {\bibinfo  {journal} {MNRAS}\ }\textbf {\bibinfo {volume} {488}},\ \bibinfo
  {pages} {4937} (\bibinfo {year} {2019})},\ \Eprint
  {http://arxiv.org/abs/1907.11268} {arXiv:1907.11268 [astro-ph.HE]}
  \BibitemShut {NoStop}%
\bibitem [{\citenamefont {{De Falco}}\ \emph {et~al.}(2020)\citenamefont {{De
  Falco}}, \citenamefont {{Bakala}},\ and\ \citenamefont
  {{Falanga}}}]{DeFalco2020}%
  \BibitemOpen
  \bibfield  {author} {\bibinfo {author} {\bibfnamefont {V.}~\bibnamefont {{De
  Falco}}}, \bibinfo {author} {\bibfnamefont {P.}~\bibnamefont {{Bakala}}}, \
  and\ \bibinfo {author} {\bibfnamefont {M.}~\bibnamefont {{Falanga}}},\ }\href
  {\doibase 10.1103/PhysRevD.101.124031} {\bibfield  {journal} {\bibinfo
  {journal} {\prd}\ }\textbf {\bibinfo {volume} {101}},\ \bibinfo {eid}
  {124031} (\bibinfo {year} {2020})},\ \Eprint
  {http://arxiv.org/abs/2006.01452} {arXiv:2006.01452 [gr-qc]} \BibitemShut
  {NoStop}%
\bibitem [{\citenamefont {{De Falco}}(2019)}]{DeFalcoTESI}%
  \BibitemOpen
  \bibfield  {author} {\bibinfo {author} {\bibfnamefont {V.}~\bibnamefont {{De
  Falco}}},\ }\href@noop {} {\bibfield  {journal} {\bibinfo  {journal} {arXiv
  e-prints}\ ,\ \bibinfo {eid} {arXiv:1904.01013}} (\bibinfo {year} {2019})},\
  \Eprint {http://arxiv.org/abs/1904.01013} {arXiv:1904.01013 [gr-qc]}
  \BibitemShut {NoStop}%
\bibitem [{\citenamefont {{De Falco}}\ and\ \citenamefont
  {{Bakala}}(2020)}]{DeFalco2019ST}%
  \BibitemOpen
  \bibfield  {author} {\bibinfo {author} {\bibfnamefont {V.}~\bibnamefont {{De
  Falco}}}\ and\ \bibinfo {author} {\bibfnamefont {P.}~\bibnamefont
  {{Bakala}}},\ }\href {\doibase 10.1103/PhysRevD.101.024025} {\bibfield
  {journal} {\bibinfo  {journal} {PRD}\ }\textbf {\bibinfo {volume} {101}},\
  \bibinfo {eid} {024025} (\bibinfo {year} {2020})},\ \Eprint
  {http://arxiv.org/abs/1911.03649} {arXiv:1911.03649 [hep-th]} \BibitemShut
  {NoStop}%
\bibitem [{\citenamefont {{De Falco}}(2020)}]{DeFalco2020sum}%
  \BibitemOpen
  \bibfield  {author} {\bibinfo {author} {\bibfnamefont {V.}~\bibnamefont {{De
  Falco}}},\ }\href@noop {} {\bibfield  {journal} {\bibinfo  {journal} {arXiv
  e-prints}\ ,\ \bibinfo {eid} {arXiv:2006.01462}} (\bibinfo {year} {2020})},\
  \Eprint {http://arxiv.org/abs/2006.01462} {arXiv:2006.01462 [gr-qc]}
  \BibitemShut {NoStop}%
\bibitem [{\citenamefont {Guckenheimer}\ and\ \citenamefont
  {Holmes}(2002)}]{Guckenheimer2002}%
  \BibitemOpen
  \bibfield  {author} {\bibinfo {author} {\bibfnamefont {J.}~\bibnamefont
  {Guckenheimer}}\ and\ \bibinfo {author} {\bibfnamefont {P.}~\bibnamefont
  {Holmes}},\ }\href {https://books.google.it/books?id=cumGDmhaFnkC} {\emph
  {\bibinfo {title} {Nonlinear Oscillations, Dynamical Systems, and
  Bifurcations of Vector Fields}}},\ Applied Mathematical Sciences\ (\bibinfo
  {publisher} {Springer New York},\ \bibinfo {year} {2002})\BibitemShut
  {NoStop}%
\bibitem [{\citenamefont {Tabor}(1989)}]{Tabor1989}%
  \BibitemOpen
  \bibfield  {author} {\bibinfo {author} {\bibfnamefont {M.}~\bibnamefont
  {Tabor}},\ }\href {https://books.google.it/books?id=5FfvAAAAMAAJ} {\emph
  {\bibinfo {title} {Chaos and integrability in nonlinear dynamics: an
  introduction}}},\ Wiley-Interscience publication\ (\bibinfo  {publisher}
  {Wiley},\ \bibinfo {year} {1989})\BibitemShut {NoStop}%
\bibitem [{\citenamefont {{De Falco}}\ \emph {et~al.}(2018)\citenamefont {{De
  Falco}}, \citenamefont {{Battista}},\ and\ \citenamefont
  {{Falanga}}}]{DeFalco2018}%
  \BibitemOpen
  \bibfield  {author} {\bibinfo {author} {\bibfnamefont {V.}~\bibnamefont {{De
  Falco}}}, \bibinfo {author} {\bibfnamefont {E.}~\bibnamefont {{Battista}}}, \
  and\ \bibinfo {author} {\bibfnamefont {M.}~\bibnamefont {{Falanga}}},\ }\href
  {\doibase 10.1103/PhysRevD.97.084048} {\bibfield  {journal} {\bibinfo
  {journal} {Physical Review D}\ }\textbf {\bibinfo {volume} {97}},\ \bibinfo
  {eid} {084048} (\bibinfo {year} {2018})},\ \Eprint
  {http://arxiv.org/abs/1804.00519} {arXiv:1804.00519 [gr-qc]} \BibitemShut
  {NoStop}%
\bibitem [{\citenamefont {De~Falco}\ and\ \citenamefont
  {Battista}(2019)}]{DeFalco2019}%
  \BibitemOpen
  \bibfield  {author} {\bibinfo {author} {\bibfnamefont {V.}~\bibnamefont
  {De~Falco}}\ and\ \bibinfo {author} {\bibfnamefont {E.}~\bibnamefont
  {Battista}},\ }\href {\doibase 10.1209/0295-5075/127/30006} {\bibfield
  {journal} {\bibinfo  {journal} {EPL}\ }\textbf {\bibinfo {volume} {127}},\
  \bibinfo {pages} {30006} (\bibinfo {year} {2019})},\ \Eprint
  {http://arxiv.org/abs/1907.13354} {arXiv:1907.13354 [gr-qc]} \BibitemShut
  {NoStop}%
\bibitem [{\citenamefont {{De Falco}}\ and\ \citenamefont
  {{Battista}}(2020)}]{DeFalco2019VE}%
  \BibitemOpen
  \bibfield  {author} {\bibinfo {author} {\bibfnamefont {V.}~\bibnamefont {{De
  Falco}}}\ and\ \bibinfo {author} {\bibfnamefont {E.}~\bibnamefont
  {{Battista}}},\ }\href {\doibase 10.1103/PhysRevD.101.064040} {\bibfield
  {journal} {\bibinfo  {journal} {Phys. Rev. D}\ }\textbf {\bibinfo {volume}
  {101}},\ \bibinfo {eid} {064040} (\bibinfo {year} {2020})},\ \Eprint
  {http://arxiv.org/abs/2003.04416} {arXiv:2003.04416 [gr-qc]} \BibitemShut
  {NoStop}%
\bibitem [{\citenamefont {Strogatz}(2000)}]{Strogatz1994}%
  \BibitemOpen
  \bibfield  {author} {\bibinfo {author} {\bibfnamefont {S.~H.}\ \bibnamefont
  {Strogatz}},\ }\href@noop {} {\emph {\bibinfo {title} {{Nonlinear Dynamics
  and Chaos: With Applications to Physics, Biology, Chemistry and
  Engineering}}}}\ (\bibinfo  {publisher} {Westview Press},\ \bibinfo {year}
  {2000})\BibitemShut {NoStop}%
\bibitem [{\citenamefont {Mori}\ \emph {et~al.}(2013)\citenamefont {Mori},
  \citenamefont {Paquette},\ and\ \citenamefont {Kuramoto}}]{Mori2013}%
  \BibitemOpen
  \bibfield  {author} {\bibinfo {author} {\bibfnamefont {H.}~\bibnamefont
  {Mori}}, \bibinfo {author} {\bibfnamefont {G.}~\bibnamefont {Paquette}}, \
  and\ \bibinfo {author} {\bibfnamefont {Y.}~\bibnamefont {Kuramoto}},\ }\href
  {https://books.google.it/books?id=lzPqCAAAQBAJ} {\emph {\bibinfo {title}
  {Dissipative Structures and Chaos}}}\ (\bibinfo  {publisher} {Springer Berlin
  Heidelberg},\ \bibinfo {year} {2013})\BibitemShut {NoStop}%
\bibitem [{\citenamefont {{Misner}}\ \emph {et~al.}(1973)\citenamefont
  {{Misner}}, \citenamefont {{Thorne}},\ and\ \citenamefont
  {{Wheeler}}}]{Misner1973}%
  \BibitemOpen
  \bibfield  {author} {\bibinfo {author} {\bibfnamefont {C.~W.}\ \bibnamefont
  {{Misner}}}, \bibinfo {author} {\bibfnamefont {K.~S.}\ \bibnamefont
  {{Thorne}}}, \ and\ \bibinfo {author} {\bibfnamefont {J.~A.}\ \bibnamefont
  {{Wheeler}}},\ }\href@noop {} {\emph {\bibinfo {title} {San Francisco:
  W.H.~Freeman and Co., 1973}}}\ (\bibinfo {year} {1973})\BibitemShut {NoStop}%
\bibitem [{\citenamefont {{Levin}}\ and\ \citenamefont
  {{Perez-Giz}}(2009)}]{Levin2009}%
  \BibitemOpen
  \bibfield  {author} {\bibinfo {author} {\bibfnamefont {J.}~\bibnamefont
  {{Levin}}}\ and\ \bibinfo {author} {\bibfnamefont {G.}~\bibnamefont
  {{Perez-Giz}}},\ }\href {\doibase 10.1103/PhysRevD.79.124013} {\bibfield
  {journal} {\bibinfo  {journal} {PRD}\ }\textbf {\bibinfo {volume} {79}},\
  \bibinfo {eid} {124013} (\bibinfo {year} {2009})},\ \Eprint
  {http://arxiv.org/abs/0811.3814} {arXiv:0811.3814 [gr-qc]} \BibitemShut
  {NoStop}%
\bibitem [{\citenamefont {{Holmes}}\ and\ \citenamefont
  {{Marsden}}(1982{\natexlab{a}})}]{Holmes1982V1}%
  \BibitemOpen
  \bibfield  {author} {\bibinfo {author} {\bibfnamefont {P.~J.}\ \bibnamefont
  {{Holmes}}}\ and\ \bibinfo {author} {\bibfnamefont {J.~E.}\ \bibnamefont
  {{Marsden}}},\ }\href {\doibase 10.1063/1.525415} {\bibfield  {journal}
  {\bibinfo  {journal} {Journal of Mathematical Physics}\ }\textbf {\bibinfo
  {volume} {23}},\ \bibinfo {pages} {669} (\bibinfo {year}
  {1982}{\natexlab{a}})}\BibitemShut {NoStop}%
\bibitem [{\citenamefont {{Holmes}}\ and\ \citenamefont
  {{Marsden}}(1982{\natexlab{b}})}]{Holmes1982V2}%
  \BibitemOpen
  \bibfield  {author} {\bibinfo {author} {\bibfnamefont {P.~J.}\ \bibnamefont
  {{Holmes}}}\ and\ \bibinfo {author} {\bibfnamefont {J.~E.}\ \bibnamefont
  {{Marsden}}},\ }\href {\doibase 10.1007/BF01961239} {\bibfield  {journal}
  {\bibinfo  {journal} {Communications in Mathematical Physics}\ }\textbf
  {\bibinfo {volume} {82}},\ \bibinfo {pages} {523} (\bibinfo {year}
  {1982}{\natexlab{b}})}\BibitemShut {NoStop}%
\bibitem [{\citenamefont {{Levin}}(1999)}]{Levin1999}%
  \BibitemOpen
  \bibfield  {author} {\bibinfo {author} {\bibfnamefont {J.}~\bibnamefont
  {{Levin}}},\ }\href {\doibase 10.1103/PhysRevD.60.064015} {\bibfield
  {journal} {\bibinfo  {journal} {\prd}\ }\textbf {\bibinfo {volume} {60}},\
  \bibinfo {eid} {064015} (\bibinfo {year} {1999})},\ \Eprint
  {http://arxiv.org/abs/astro-ph/9811213} {arXiv:astro-ph/9811213 [astro-ph]}
  \BibitemShut {NoStop}%
\end{thebibliography}%
\end{document}